\documentclass[aps,pra,twocolumn,superscriptaddress,nobalancelastpage,longbibliography]{revtex4-1}

\usepackage{amsmath,amssymb,mathtools}
\usepackage[dvipdfmx]{graphicx}
\usepackage[colorlinks=true,bookmarks=true,citecolor=blue,linkcolor=blue,urlcolor=blue,breaklinks=true]{hyperref}

\usepackage{dcolumn}
\usepackage{bm}
\usepackage[T1]{fontenc}
\usepackage{textcomp}
\usepackage{color}
\usepackage{mathrsfs}
\usepackage[normalem]{ulem}
\usepackage{enumerate}
\usepackage{setspace}
\usepackage{physics}

\renewcommand\({\begin{equation}}	
\renewcommand\){\end{equation}}

\renewcommand\[{\begin{eqnarray}}	
\renewcommand\]{\end{eqnarray}}
\newcommand{\al}[1]{\begin{aligned}#1\end{aligned}}

\begin{document}

\title{Magnonic $\varphi$ Josephson junctions
and synchronized precession}

\author{Kouki Nakata}
\affiliation{Advanced Science Research Center, Japan Atomic Energy Agency, Tokai, Ibaraki 319-1195, Japan}

\author{Ji Zou}
\affiliation{Department of Physics, University of Basel, Klingelbergstrasse 82, 4056 Basel, Switzerland}

\author{Jelena Klinovaja}
\affiliation{Department of Physics, University of Basel, Klingelbergstrasse 82, 4056 Basel, Switzerland}

\author{Daniel Loss}
\affiliation{Department of Physics, University of Basel, Klingelbergstrasse 82, 4056 Basel, Switzerland}
\affiliation{Center for Emergent Matter Science, RIKEN, 2-1 Hirosawa, Wako-shi, Saitama 351-0198, Japan}

\date{\today}

\begin{abstract}
There has been a growing interest in non-Hermitian physics. One of its main goals is to engineer dissipation and to explore ensuing functionality. In magnonics, the effect of dissipation due to local damping on magnon transport has been explored. However, the effects of non-local damping on the magnonic analog of the Josephson effect 
remain missing, despite that non-local damping is inevitable and has been playing a central role in magnonics.  
Here, we uncover theoretically that a surprisingly rich dynamics can emerge in magnetic junctions  due to intrinsic non-local damping, using analytical and numerical methods. In particular, under microwave pumping, we show that coherent spin precession in the right and left insulating ferromagnet (FM) of the junction becomes synchronized
by non-local damping and thereby a magnonic analog of the $\varphi$ Josephson junction emerges, where $\varphi$ stands here for the relative precession phase of right and left FM in the stationary limit. Remarkably,  $\varphi$ decreases monotonically from  $ \pi$ to $\pi/2$ as the magnon-magnon interaction, arising from spin anisotropies,
increases.  Moreover, we also find a magnonic  diode effect giving rise to rectification  of  magnon currents. Our predictions are readily testable with  current device and measurement technologies at room temperatures.
\end{abstract}

\maketitle


\section{Introduction}
\label{sec:I}

Recently,  non-Hermitian physics~\cite{Review_NH_Ashida}  has been attracting growing interest from both 
fundamental science  and  applications  such as energy-efficient devices.
One of its main themes is to engineer dissipation and to explore the resulting functionality.
Since magnons are intrinsically damped in magnetic systems, the goal of non-Hermitian physics aligns well with
magnonics~\cite{MagnonSpintronics,Chumak_Roadmap_SWcomputing,Review_QuantumMagnonics},
which aims at  efficient transmission and processing of information  for computing and communication technologies
using magnons as its carrier in units of the Bohr magneton $\mu_{\text{B}}$.
To this end, establishing methods for the control and manipulation of  magnon transport subjected to dissipation
is crucial. In magnonics, the effect of dissipation due to local (Gilbert) damping  on magnon transport,
such as the magnonic analog of the Josephson effect~\cite{KKPD},
has been explored~\cite{TroncosoJosephson,AFJosephsonTHz,AFJosephsonTHz2},
where the macroscopic coherent magnon state,
the key ingredient for the magnonic Josephson effect, 
realizes an oscillating behavior of magnon transport. However, the effect of non-local damping on the magnonic Josephson effect and the resulting functionality remain missing, despite that  non-local damping is inevitable and has been playing a central role  in non-Hermitian magnonics~\cite{ReviewNHmagnonics,YT_EP_LLG,Ji_NHdiode}.

\begin{figure}[!b]
\begin{center}
\includegraphics[width=8.8cm,clip]{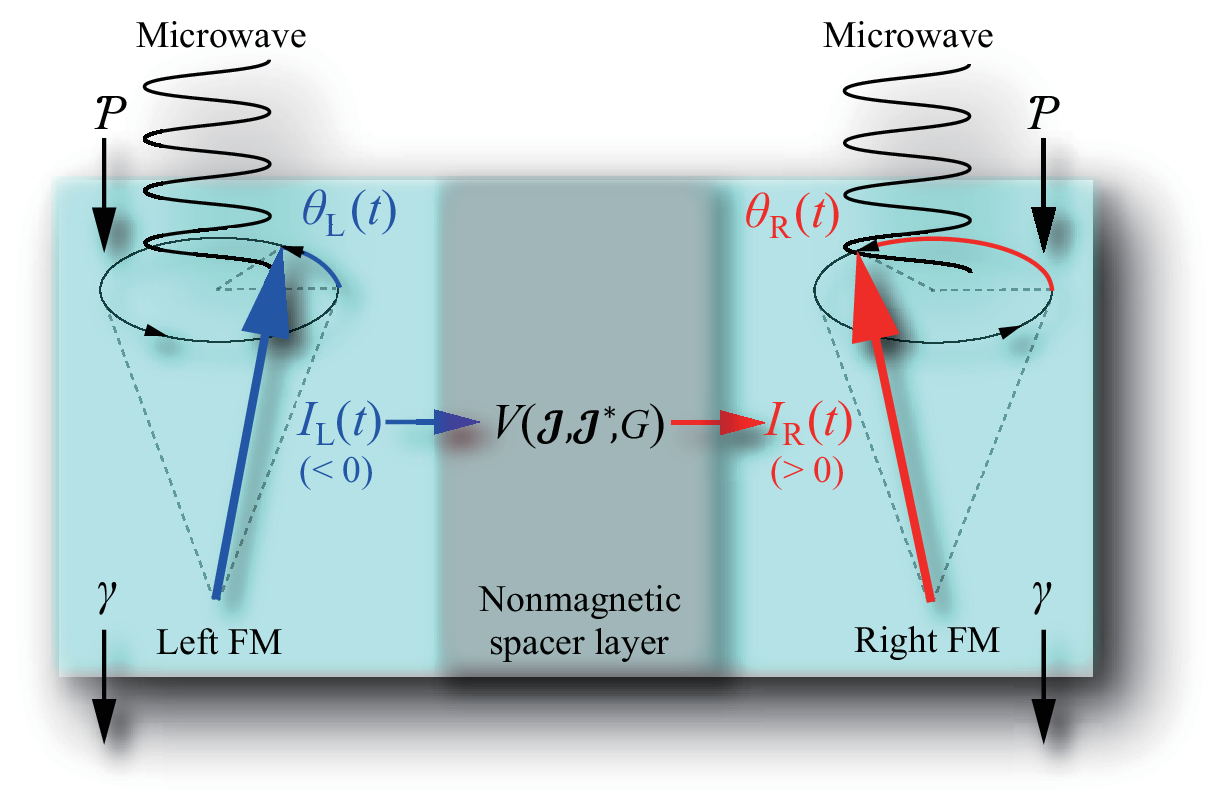}
\caption{Magnonic $\varphi$ Josephson junction, 
formed by a ferromagnetic bilayer coupled by a non-magnetic spacer. 
Magnons are subjected to local damping
at rate $ \gamma > 0 $.
The  spacer layer-mediated interaction $V$ between the two FMs consists of coherent coupling (${\mathcal{J}}$) and non-local damping ($G$). Under microwave pumping, the  spatially uniform mode of magnons is injected into each FM
at  rate $\mathcal{P}$, generating coherent spin precessions. The loss of magnons due to dissipation 
is exactly balanced by the injection of magnons. In the steady state,  the  spin precessions in each FM 
get synchronized due to non-local damping and a relative precession angle emerges
$ \varphi =\lim_{t \to \infty}  
[\theta_{\text{R}}(t)-\theta_{\text{L}}(t)]  $
that depends on the magnon-magnon interaction. The associated magnon currents $I_{{\text{L}},{\text{R}}}$ get
rectified when $G\neq 0$.}
\label{fig:1}
\end{center}
\end{figure}

Here, we fill this gap. 
We find that the inherent non-local damping leads to rich dynamics, and, interestingly,  can be utilized to realize
a magnonic analog of the $\varphi$ Josephson junction~\cite{piTheory1977,piTheory1987,piTheory1992,piExp1993,piExp1995,piExp2001,phiExp2001,phiExp2004,phiExp2012,phiExp2013,phiExp2016,phiTheory1996,phiTheory1996_2,phiTheory1997,phiTheory1998,phiTheory2001,phiTheory2003,phiTheory2007,PhiTheory2008,phiTheory2011}~\footnote{See Ref.~\cite{AJE_exp_2023} for observation of anomalous Josephson effect in metallic and nonmagnetic superconducting-normal-superconducting junctions configured as nonequilibrium Andreev interferometers~\cite{AJE_Theory_2019,AJE_Theory_2019PRB,AJE_Theory_2021}.}. 
Under microwave pumping, coherent spin precession  in each insulating ferromagnet (FM)
of the magnetic junction (see Fig.~{\ref{fig:1}}) is synchronized  by non-local damping as time advances
and gives rise of a magnonic $\varphi$ Josephson junction, where $\varphi$ stands for the stationary value of the relative precession phase of the left and right FMs.
Interestingly, we find that  $\varphi$ decreases monotonically from $\pi$ to $\pi/2$ as the magnon-magnon interaction, arising from spin anisotropies across the junction, increases.
Applying microwaves to each FM continuously, the junction reaches the nonequilibrium steady state where 
the loss of magnons due to dissipation is precisely balanced  by the injection of magnons. 
Hence, spins in each FM continue to precess coherently, and the synchronized precession  of the left and the right magnetization remains stable even at room temperature. Finally, we show that  the magnetic junction exhibits
rectification and acts as a diode for the magnon current.

This paper is organized as follows.
In Sec.~\ref{sec:II},
we introduce the model system for coherent magnons
under microwave pumping,
magnonic Josephson junction,
and derive the magnonic Josephson equations 
in the presence of non-local damping.
We then describe the synchronized precession 
of the left and the right magnetization
in magnonic $\varphi$ Josephson junctions 
and study the behavior of $ \varphi$ 
as a function of magnon-magnon interactions
in Sec.~\ref{sec:III}.
We also show that 
magnonic $\varphi$ Josephson junctions
exhibit a rectification due to non-local damping
in Sec.~\ref{sec:IV}.
In Sec.~\ref{sec:VI},
we discuss experimental feasibility of 
our theoretical predictions.
Finally, 
we summarize and give some conclusions in Sec.~\ref{sec:VII}.
Technical details are deferred to the Appendices.

\section{Magnonic Josephson junctions}
\label{sec:II}

We consider a magnetic junction, shown in Fig.~{\ref{fig:1}}, consisting of
a bilayer of two insulating ferromagnetic layers, where the two  FMs  are separated by a nonmagnetic spacer layer, thereby weakly spin-exchange coupled. Applying microwaves to each FM and tuning the microwave frequency $\Omega >0$ of GHz to the magnon energy gap, ferromagnetic resonance is generated for each FM separately,
where spins precess coherently. Under this microwave pumping, the zero wavenumber mode (i.e., spatially uniform mode) of magnons is excited and injected into each FM at the rate of $\mathcal{P}>0$~\cite{KPD}. The magnons are subjected to local Gilbert damping at the rate  $ \gamma > 0 $ in each FM.

In addition, there is non-local damping~\cite{ReviewNHmagnonics,YT_EP_LLG} that is mediated by the  spacer  interaction between the two FMs, see Fig.~{\ref{fig:1}}. Using the Holstein-Primakoff transformation~\cite{HP}, this coupling term becomes to leading order~\cite{Ji_NHdiode}
$ V= 
({\mathcal{J}}-i\hbar G/2)
a_{\text{L}}^{\dagger}a_{\text{R}} 
+
({\mathcal{J}}^{*}-i\hbar G/2)
a_{\text{L}}a_{\text{R}}^{\dagger}$,
where $  a_{\text{L(R)}}^{(\dagger)}  $ represents the magnon annihilation (creation) operator
for the zero wavenumber mode, which satisfies the bosonic commutation relation. Here, $ {\mathcal{J}} \in {\mathbb{C}}$, with its complex conjugate ${\mathcal{J}}^{*} $, describes coherent coupling,
while $\hbar G \in {\mathbb{R}}$ describes the non-local damping (or dissipative coupling)~\cite{Ji_NHdiode}.
The real part of coherent coupling, ${\text{Re}}({\mathcal{J}})= {\mathcal{J}}_{\text{Re}}$,
arises from  the symmetric spin-exchange interaction between the two FMs,
while its imaginary part~\cite{KKPD,KPD,magnon2,ACphaseExp2023},
${\text{Im}}({\mathcal{J}})= {\mathcal{J}}_{\text{Im}} $,
is induced, for example, by the Dzyaloshinskii-Moriya interaction when the spatial inversion symmetry of 
the nonmagnetic spacer layer is broken~\cite{katsura2}. 
Each component of $ {\mathcal{J}}/\hbar $ can reach MHz in experiments~\cite{ImJ2023MHz,ImJ2023MHz2}.
The values of  $\gamma $ and $ G $ are typically within the MHz regime~\cite{Damping_Heinrich2003},
and the condition, $ \hbar \Omega   \gg  |\mathcal{J}|, \hbar \gamma, |\hbar G|  $, is satisfied. The condition 
$ |G| \leq 2 \gamma  $
ensures the complete positivity of the system dynamics~\cite{Ji_Dissipative_Bell,Ji_NHdiode,zou2023spatially},
and we focus on this regime henceforth. We note that the coupling term $V$ 
becomes non-Hermitian due to non-local damping, i.e., $  V\neq V^{\dagger} $ for $  G\neq 0 $.

The magnon-magnon interaction in the left (right) FM,
$ U_{\text{L(R)}}  $, arises from anisotropies of spin, where
$ U_{\text{L}}
= ({U}/{2}) 
a_{\rm{L}}^{\dagger}a_{\rm{L}}^{\dagger}
a_{\rm{L}}a_{\rm{L}}  $,
$U_{\text{R}}
= ({U}/{2}) 
a_{\text{R}}^{\dagger}a_{\text{R}}^{\dagger}
a_{\text{R}}a_{\text{R}}$.  Here, $U$ can take 
both positive and negative values, depending on the combination of anisotropies
such as  the spin anisotropy along the quantization axis 
and the anisotropy of the spin-exchange interaction in each FM~\cite{KKPD,KPD}.

In this study, we envisage to continuously apply microwaves to each FM, which results in spins exhibiting macroscopic coherent  precession characterized as $  \langle a_{\text{L(R)}}(t) \rangle \neq  0 $~\cite{KPD}.
Therefore, assuming a macroscopic coherent magnon state,
thereby using the semiclassical approximation, we replace the operators $ a_{\text{L}}(t) $
and $ a_{\text{R}}(t) $ by their expectation values as
$ \langle a_{\text{L}}(t) \rangle 
=\sqrt{N_{\text{L}}(t)} 
{\text{e}}^{i\theta_{\text{L}}(t)}  $
and
$ \langle a_{\text{R}}(t) \rangle
=\sqrt{N_{\text{R}}(t)} 
{\text{e}}^{i\theta_{\text{R}}(t)}  $,
respectively,
where $ N_{\text{L(R)}}(t) >0 $
represents the number of coherent magnons
for each site in the left (right) FM and  $ \theta_{\text{L(R)}} $ is the phase (see Fig.~{\ref{fig:1}}). 
Defining the relative precession phase as
$ \theta(t)= \theta_{\text{R}}(t)-\theta_{\text{L}}(t) $,
each time evolution
[e.g., $  {\dot{\theta}}(t) $ represents
the time derivative of $ \theta(t) $]
is described as (see the Appendices for details \cite{NHJosephsonSM})
\begin{widetext}
\begin{subequations}
\begin{align}
&  {\dot{\theta}}(t)
= 
\frac{{\mathcal{J}}_{\text{Re}}}{\hbar}
\cos \theta(t)
\left(
\sqrt{\frac{N_{\text{R}}(t)}{N_{\text{L}}(t)}}
-\sqrt{\frac{N_{\text{L}}(t)}{N_{\text{R}}(t)}}
\right)
+\sin \theta(t)
\left[
\left(\frac{{\mathcal{J}}_{\text{Im}}}{\hbar}
+\frac{G}{2}\right)
\sqrt{\frac{N_{\text{L}}(t)}{N_{\text{R}}(t)}}
-\left(\frac{{\mathcal{J}}_{\text{Im}}}{\hbar}
-\frac{G}{2}\right) 
\sqrt{\frac{N_{\text{R}}(t)}{N_{\text{L}}(t)}}
\right]
+u[N_{\text{L}}(t)-N_{\text{R}}(t)],
\label{eqn:4a}  \\
&{\dot{N}}_{\text{L/R}}(t)
= -2 \gamma N_{\text{L/R}}(t)
\pm 2\left[
\frac{{\mathcal{J}}_{\text{Re}}}{\hbar}
\sin \theta(t)
+\left(\frac{{\mathcal{J}}_{\text{Im}}}{\hbar}
\mp \frac{G}{2}\right) 
\cos \theta(t)
\right]
\sqrt{N_{\text{L}}(t)N_{\text{R}}(t)}
+{\mathcal{P}},
\label{eqn:4b}  
\end{align}
\end{subequations}
\end{widetext}
where
$u=U/\hbar$.
The second term 
on the right-hand side of 
Eq.~(\ref{eqn:4b})
describes the nonmagnetic spacer layer-mediated transport of 
coherent magnons in the junction.

The transport of coherent magnons in this junction
is analogous to the Josephson effect~\cite{Josephson1962}
in the sense that  the current arises  as the term of the order of $V$,
$ O(V) $, for the interaction $V=V({\mathcal{J}}, {\mathcal{J}}^{*}, G)$ 
between the two FMs (see Fig.~\ref{fig:1})
and is characterized  by the relative precession phase $\theta(t)$.
For these reasons~\cite{katsura2}, we refer to  the transport of coherent magnons
and the junction as the magnonic Josephson effect and the magnonic Josephson junction, respectively.
Note that, in contrast, the current for incoherent magnons arises from a term of $O(V^2)$~\cite{magnonWF,ReviewMagnon}.

\begin{figure}[t]
\begin{center}
\includegraphics[width=5.5cm,clip]{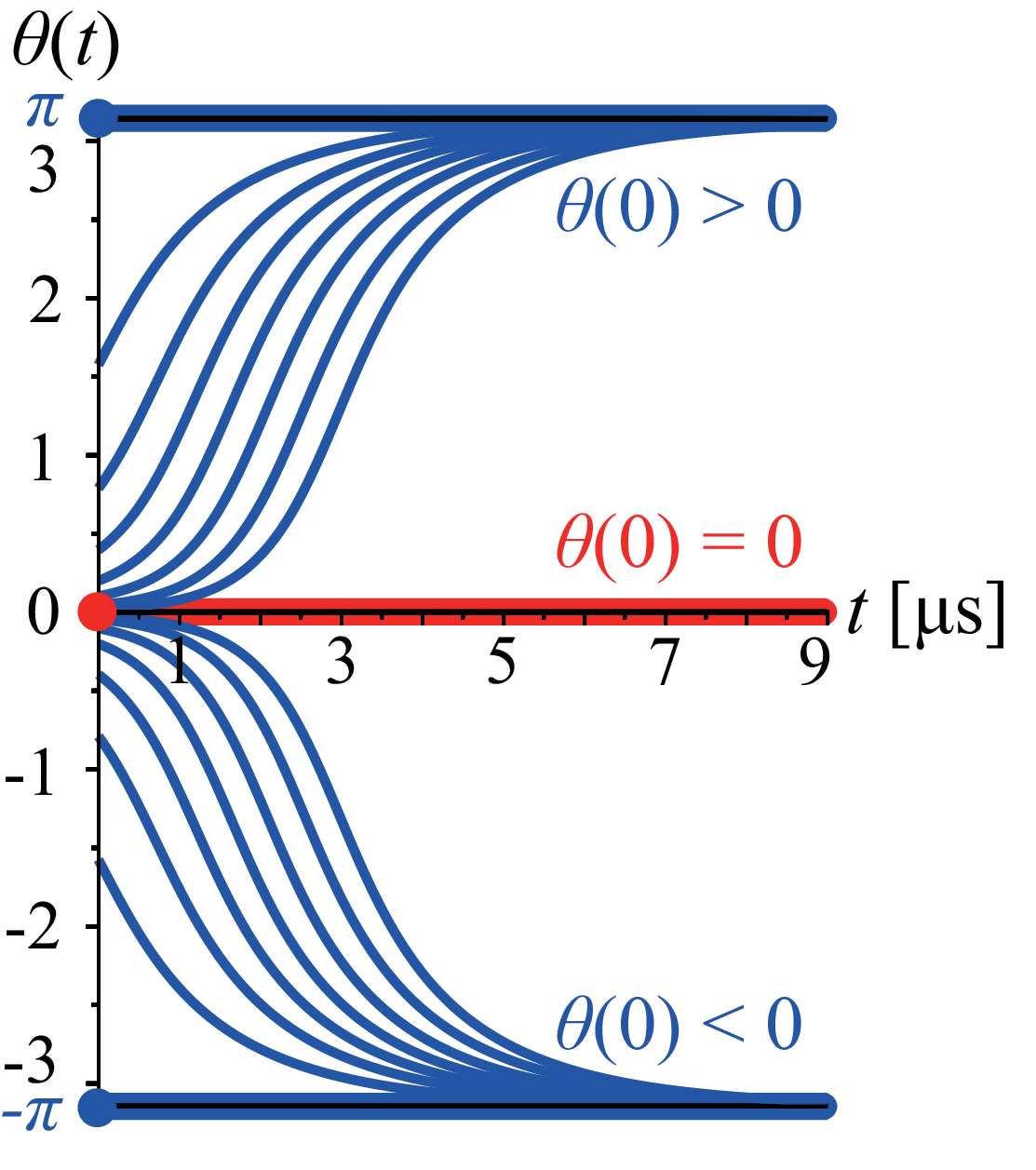}
\caption{Plots of  the relative phase $\theta(t)$ as a function of time 
in the absence of the magnon-magnon interaction,  i.e., $u=0$,
for the initial condition $ \theta(0)=0 $
and $ |\theta(0)|= \pi/2^n $ with $n=0,1,...,7$ obtained by numerically solving 
Eqs.~\eqref{eqn:10a}-\eqref{eqn:10c} for the parameter values
$ {\mathcal{J}}_{\text{Re}}=0$,
$ {\mathcal{J}}_{\text{Im}}/\hbar= G/2= 0.5$ MHz,
$ \gamma =1.1 $ MHz, $  {\mathcal{P}}= 2 \gamma =2.2  $ MHz, and $ N_{\text{L}}(0)=N_{\text{R}}(0) = 10^{-6}  $.
As time advances, coherent spin precession in each FM is synchronized  and forms the magnonic $\pi$ Josephson junction as $ \varphi = \theta(t\to \infty)= \pm \pi $ for $\theta (0) \neq  0 $.
The initial condition $\theta(0)=0$ for $u=0$  results in $ \varphi = 0 $.}
\label{fig:2}
\end{center}
\end{figure}

\section{Synchronized precession}
\label{sec:III}

To seek for a junction setup that exhibits synchronization and rectification effects,
we consider the case where
\begin{equation}
{\mathcal{J}}_{\text{Re}}  =0
\  \
\text{and}
\  \
 {\mathcal{J}}_{\text{Im}}/\hbar = G/2 >0.
\label{eqn:9}
\end{equation}
Under these assumptions, Eqs.~\eqref{eqn:4a}-\eqref{eqn:4b} become
\begin{subequations}
\begin{align}
&  {\dot{\theta}}(t)
= 
G 
\sqrt{N_{\text{L}}/N_{\text{R}}}
\sin \theta
+u(N_{\text{L}}-N_{\text{R}}),
\label{eqn:10a}  \\
&{\dot{N}}_{\text{L}}
= -2 \gamma N_{\text{L}}
+{\mathcal{P}},
\label{eqn:10b}  \\
&{\dot{N}}_{\text{R}}
= -2 \gamma  N_{\text{R}}
-2G
\sqrt{N_{\text{L}}N_{\text{R}}}
\cos \theta
+{\mathcal{P}},
\label{eqn:10c} 
\end{align}
\end{subequations}
where we suppressed for brevity the explicit time-dependence of the quantities $\theta(t)$ and $N_{\text{L/R}}(t)$.
Under microwave pumping $\mathcal{P}$,
the nonequilibrium steady state 
$ {\dot{\theta}}(t)={\dot{N}}_{\text{L}}(t)= {\dot{N}}_{\text{R}}(t)=0  $ is realized, where $\theta(t) $, $ N_{\text{L}}(t)$, and $ N_{\text{R}}(t) $ approach their stationary values as time advances,  i.e., $\varphi=\theta(t \to \infty)$  and $N_{\text{L(R)}}(t \to \infty)  =   N_{\text{L(R)}}^{\infty} $. Using Eqs.~\eqref{eqn:10a}-\eqref{eqn:10c}, we find the following relations between these asymptotic quantities:
\begin{subequations}
\begin{align}
&\cos \varphi = (\gamma/G) (N_{\text{L}}^{\infty}-N_{\text{R}}^{\infty})/\sqrt{N_{\text{R} }^{\infty}N_{\text{L}}^{\infty}} ,
\label{eqn:NRinf}\\
&\tan \varphi = - u N_{\text{R}}^{\infty}/ {\gamma},
 \label{eqn:selfconsistent}
\end{align}
\end{subequations}
where $N_{\text{L}}^{\infty}  = 
\mathcal{P}/2 \gamma$ is a constant, while $N_{\text{R}}^{\infty}$ becomes a function of the  relative precession phase $\varphi$.
In what follows, we will show that there is a unique solution to the system of  Eqs. (\ref{eqn:NRinf}) and (\ref{eqn:selfconsistent}).
Thereby, a magnonic analog of the $\varphi$ Josephson junction~\cite{piTheory1977,piTheory1987,piTheory1992,piExp1993,piExp1995,piExp2001,phiExp2001,phiExp2004,phiExp2012,phiExp2013,phiExp2016,phiTheory1996,phiTheory1996_2,phiTheory1997,phiTheory1998,phiTheory2001,phiTheory2003,phiTheory2007,PhiTheory2008,phiTheory2011}
is realized.

\begin{figure*}[t]
\begin{center}
\includegraphics[width=18cm,clip]{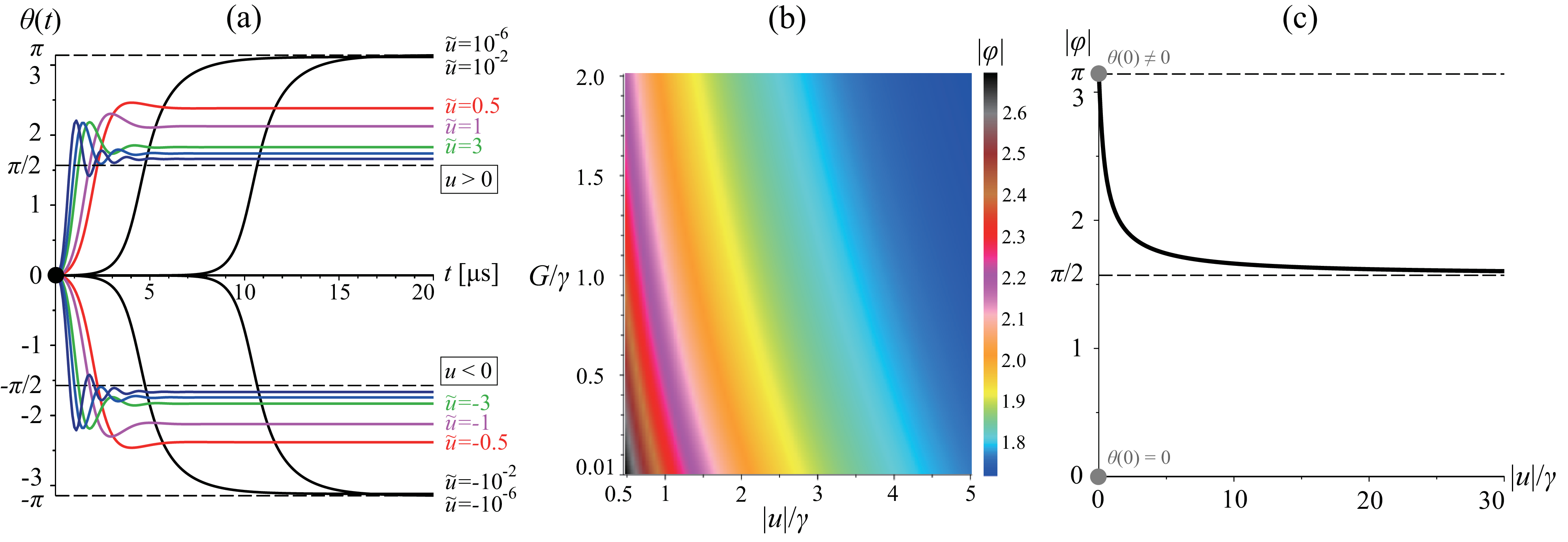}
\caption{
(a) 
Plots of  the relative phase $\theta(t)$ as a function of time  for several values of $\tilde u$ 
obtained by numerically solving Eqs.~\eqref{eqn:10a}-\eqref{eqn:10c}
for the same parameter values as in Fig.~\ref{fig:2} as well as for $\theta(0)=0$.
The stationary relative  phase, $\varphi= \theta(t \to \infty) $, decreases from  $ |\varphi| =\pi$ to  $ |\varphi| =\pi/2$
as the magnon-magnon interaction $|\tilde{u}|$ increases, where we set $ |\tilde{u}|=10^{-6},10^{-2},0.5,1,3,5, 10$.
We note that $ {\text{sgn}}(\varphi) = {\text{sgn}}(u) $. (b,c) The relative phase $|\varphi|$ decreases as $ |u|/\gamma $ and $ G/\gamma $ increase. These plots are obtained by numerically solving Eq.~\eqref{eqn:varphiEq} for the same parameter values as in Fig.~\ref{fig:3}(a).
(c) Plot of $|\varphi|$ as a function of $ |u|/\gamma $, showing that $|\varphi|$ decreases monotonically from 
$\pi$ to $\pi/2$ as $|u|/\gamma$ increases. For $u\neq0$, the value of $|\varphi|$  does not depend on the initial conditions. For $u = 0$, $|\varphi|$ depends on the initial condition $\theta(0)$ as $|\varphi|=\pi$ for $\theta(0)\neq0$,
whereas $\varphi=0$ for $\theta(0)=0$ (see Fig.~\ref{fig:2}).}
\label{fig:3}
\end{center}
\end{figure*}

\subsection{Absence of magnon interaction: $u=0$}
\label{subsec:III-1}

First, we study the behavior of $ \varphi$ in the absence of the magnon-magnon interaction, $u=0$.  From Eq.~\eqref{eqn:selfconsistent}, we immediately find that $\varphi$ can only be equal to $\pm \pi$ or $0$. One of these three values is chosen based on the initial condition $\theta(0)$. Figure~\ref{fig:2} shows the plots of  the relative phase $\theta(t)$ for several initial conditions $\theta(0)$ as a function of time  obtained by numerically solving Eqs.~\eqref{eqn:10a}-\eqref{eqn:10c}. If $ \theta (0) =0$, the relative phase stays constant, $\theta (t) =0 $. If the symmetry is broken, i.e.,  $ \theta (0) \neq 0$, we have $\varphi = \pi\ \rm{sgn}\  \theta (0)$. This shows that the
coherent spin precessions in each FM get synchronized with each other as time advances. This asymptotic locking of the spin precessions is a direct consequence of the dissipative coupling term $G$. The junction behavior represents a magnonic analog of  the well-known $\pi$ Josephson junction effect in superconductors~\cite{piTheory1977,piTheory1987,piTheory1992,piExp1993,piExp1995,piExp2001}.

Figure~\ref{fig:2} also shows that  the point $\theta(0)=0$  is unstable in the sense that  $ \varphi = 0 $ 
for  the initial condition $ \theta (0)=0  $, whereas $ \varphi = \pm \pi $ for $ \theta (0) \neq  0 $. However, to 
realize such a special condition,
$\theta(0)=0$ with $u=0$, 
will be out of experimental reach. Moreover, in what follows, we will show that any finite value of $u$ results in $\varphi\neq0$, no matter what initial value we choose for 
$\theta(0)$ including the fine-tuned value $\theta(0)=0$, see Fig.~\ref{fig:3}.

We emphasize that our results are independent of the initial values
$N_{\text{L}}(0)$ and $N_{\text{R}}(0)$
and do not depend on the assumption that both FMs are pumped at the same rate $\mathcal{P}$. A detailed discussion is available in the Appendices~\cite{NHJosephsonSM}.
We also remark that 
the synchronized precession of 
the left and the right magnetization, 
$ {\dot{\theta}}(t) =0$, 
remains valid
even if the parameter values deviate 
from Eq.~\eqref{eqn:9}.
For details of the parameter dependence of our results,
see the Appendices~\cite{NHJosephsonSM},
where we numerically show
the robustness of the synchronized precession.
We note that 
under the initial condition $ \theta(0)=0 $,
we have
$ \varphi \neq 0$ for ${\mathcal{J}}_{\text{Re}}\neq 0$,
whereas
$ \varphi = 0$ for ${\mathcal{J}}_{\text{Re}}=0$.
Also in this sense, the point $ \theta(0)=0 $ is unstable.
See the Appendices~\cite{NHJosephsonSM} for more details.

\subsection{Finite magnon interaction: $u\neq0$}
\label{subsec:III-2}

Next, we study the behavior of $ \varphi$ in the presence of the magnon-magnon interaction, $u\neq0$,
and determine $ \varphi$ as a function of $u$. First, from Eq. (\ref{eqn:selfconsistent}), we deduce that $ \varphi(u)$ is an odd function of $u$, i.e.,  $ \varphi(u)=-\varphi(-u)$ (mod $2\pi$). Here, we used that $ N_{\text{R}}^{\infty}(\varphi) = N_{\text{R}}^{\infty}(-\varphi) $, which follows from Eq.~\eqref{eqn:NRinf}. In Fig. \ref{fig:3}, we solve numerically
Eqs.~\eqref{eqn:10a}-\eqref{eqn:10c} and  confirm that $ \varphi(u)$  is odd. Remarkably, even if the initial condition is chosen as $\theta(0)=0$, for $u\neq0$, the magnitude of $\varphi$ monotonically decreases from 
$ |\varphi| =\pi$ to $ |\varphi| =\pi/2$ as the magnitude of the magnon-magnon interaction increases. We note that this condition is consistent with the requirement  that the  direction of magnon propagation in the junction, see Fig.~{\ref{fig:1}}, is chosen to be from left to right (see also below).

Introducing the rescaled magnon-magnon interaction  $ {\tilde{u}}=(u/\gamma) N_{\text{L}}^{\infty}$   and combining Eqs.~\eqref{eqn:10a}-\eqref{eqn:10c}, we get the following implicit equation on $\tan \varphi$:
\begin{align}
&\tan^4\varphi
+2 \tilde{u} \tan^3\varphi
+(1+ {\tilde{u}}^2) \tan^2\varphi \nonumber \\
&\hspace{60pt}+\left[2+\left( G/\gamma \right)^2 \right]
 {\tilde{u}} \tan \varphi
+ {\tilde{u}}^2=0,
\label{eqn:varphiEq} 
\end{align}
whose solution gives us $\varphi$ as function of $ {\tilde{u}} $.
For $\varphi \in [-\pi, \pi]$, there are generally four solutions for $\varphi$. The two solutions with $\cos \varphi >0$ are excluded as they will result in an unphysical direction of the current $I_{\text{R}}$ that is set by the nonlocal interaction term $V$ (see below).  Finally, one more solution is eliminated as it breaks the condition $\tan \varphi < - \tilde{u}$, which follows from Eqs. (\ref{eqn:NRinf}) and (\ref{eqn:selfconsistent}) for $ \tilde{u}>0$. That leaves us with a unique solution for $\varphi$.

In Fig.~\ref{fig:3}(b), we investigate $|\varphi|$ as a function of both $ |u|/\gamma $ and $ G/\gamma $ by numerically  solving  Eq.~\eqref{eqn:varphiEq}. For a given $\tilde u$ ($G$),  $|\varphi|$  decreases monotonically from 
$ \pi$ to  $ \pi/2$ as $G$ ($\tilde u$) increases, see Fig.~\ref{fig:3}(b,c). The asymptotic expressions for $\varphi$ for the two limiting cases of weak and strong interactions can be obtained also analytically. Indeed, from Eqs.~\eqref{eqn:varphiEq} and~\eqref{eqn:NRinf}
and for strong interactions, $\tilde{u} \gg 1$, we get in leading order in $1/{\tilde{u}}$  
\begin{subequations}
\begin{align}
&\varphi   \simeq \pi/2+ 1/\tilde{u},
\label{eqn:uLarge}  \\
&N_{\text{R}}^{\infty}/N_{\text{L}}^{\infty}  \simeq1 +G/\tilde u \gamma.
\label{eqn:uLargeNR} 
\end{align}
\end{subequations}
In the opposite limit of weak interactions,  $ 0 < \tilde{u}  \ll 1$, we obtain, keeping leading corrections in $\tilde{u} $ in each quantity,
\begin{subequations}
\begin{align}
&\varphi 
  \simeq
\pi - c {\tilde{u}},
\label{eqn:uSmall} \\
&N_{\text{R}}^{\infty}  /  N_{\text{L}}^{\infty}  \simeq  c 
\left ( 1 -  G (c \tilde u)^2 /  \sqrt{G^2 + 4 \gamma^2 } \right )
\label{eqn:uSmallA}
\end{align}
\end{subequations}
where $c = (\sqrt{1 + \left(G/ 2\gamma\right)^2}+G/2\gamma )^2$. These equations together with Fig.~\ref{fig:3}(c) 
are one of the main results of this study. Remarkably, both limiting values for $\varphi$, namely $\pi/2$ and $\pi$, are obviously universal, i.e., independent of any material properties as well as of initial conditions. This underlines the robust and universal property of the synchronization of precessions brought about by the nonlocal damping in magnetic junctions driven by microwaves.

\section{Rectification}
\label{sec:IV}

The magnonic $\varphi$ Josephson junction can be regarded as a magnonic analog of the Josephson diode~\cite{JDtheory,JDtheoryLiangFu,JDexp}, 
in the sense that it exhibits a rectification effect for the magnon currents, as we will show next.
From Eq.~\eqref{eqn:4b} one gets that the current of coherent magnons  that flows from  the nonmagnetic spacer layer
into the left (right) FM (see Fig.~{\ref{fig:1}}),  $ I_{\text{L(R)}}(t)=O(V) $, is given by
\begin{align}
I_{\text{L/R}}
=&  \pm 2\Big[ \frac{{\mathcal{J}}_{\text{Re}}}{\hbar} \sin \theta+\Big(\frac{{\mathcal{J}}_{\text{Im}}}{\hbar}
\mp\frac{G}{2}\Big)  \cos \theta \Big] \sqrt{N_{\text{L}}N_{\text{R}}},
\label{eqn:5a} 
\end{align}
in units of $ g \mu_{\text{B}} $ for the $g$-factor $g$ of the constituent spins, and where we suppressed for brevity the explicit time-dependence of the quantities $\theta(t)$ and $N_{\text{L/R}}(t)$.
The condition specified in  Eq.~\eqref{eqn:9}
results in
\begin{subequations}
\begin{align}
I_{\text{L}}(t)=&  0,
\label{eqn:11a}  \\
I_{\text{R}}(t)=& -2G  \sqrt{N_{\text{L}}(t)N_{\text{R}}(t)} \cos \theta.
\label{eqn:11b} 
\end{align}
\end{subequations}
Since the nonequilibrium steady state  is realized under microwave pumping $\mathcal{P}$, the current  $  I_{\text{R}}(t)  $ continues to flow while keeping the rectification effect characterized by $  I_{\text{L}}(t) =0 $
in the magnonic $\varphi$ Josephson junction (see Fig.~\ref{fig:1}). We emphasize that the rectification of the magnon current holds  also in the presence of the magnon-magnon interaction.

Note that our analytical solutions for $\varphi$ must satisfy $\cos \varphi <0 $ and thus the magnitude of $\varphi$ is bounded as $ {\pi}/{2}<| \varphi | \leq \pi  $ to ensure $ \lim_{t \to \infty} I_{\text{R}}(t) \geq 0 $.
We recall that  the non-local damping $G$ provides the non-Hermitian property to the nonmagnetic spacer layer-mediated interaction described by $ V$. Under the special conditions stated in Eq.~(\ref{eqn:9}) one gets $ V= 
-i\hbar G a_{\text{L}}a_{\text{R}}^{\dagger}$. This shows that  propagation of magnons becomes chiral and is allowed only in the direction from left to right in the junction. This is consistent with the sign choice of the current and its direction as shown in Fig.~{\ref{fig:1}}.

In the absence of non-local damping, $G=0$, the current of coherent magnons propagates in both directions through the nonmagnetic spacer layer, from the left to the right FM and vice versa~\cite{KKPD}. This results in 
$  I_{\text{R}}(t)= - I_{\text{L}}(t) $ for $G=0$ [see Eqs.~(\ref{eqn:5a})]. Thus,  we see that rectification only occurs in the presence of non-local damping when $G\neq0$.

If instead of Eq.~(\ref{eqn:9}), one uses the condition
\begin{equation}
{\mathcal{J}}_{\text{Re}}  =0 \  \ \text{and} \  \ {\mathcal{J}}_{\text{Im}}/\hbar = -G/2,
\label{eqn:13}
\end{equation}
we get from Eq.~(\ref{eqn:5a})  that
\begin{subequations}
\begin{align}
I_{\text{L}}(t)
=& 
-2G  
\sqrt{N_{\text{L}}(t)N_{\text{R}}(t)}
\cos \theta(t),
\label{eqn:14a}  \\
I_{\text{R}}(t)
=& 
0.
\label{eqn:14b} 
\end{align}
\end{subequations}
Thus, the rectification effect changes its direction. We conclude that the polarity of the rectification effect is determined by the sign of $G$.

\section{Experimental feasibility}
\label{sec:VI}

The key ingredient for the magnonic Josephson effect is the coherent magnon state $  \langle a_{\text{L(R)}}(t) \rangle \neq  0 $ (i.e., coherent spin precession), and it can be realized through microwave pumping~\cite{KPD}.
Since each component of coherent coupling $ {\mathcal{J}} $ is tunable by adjusting the thickness of 
the nonmagnetic spacer layer~\cite{ParkinPRL1990,ParkinPRL1991,HusainPRB2022,yun2023anisotropic,ImJ2023MHz}
or applying an electric field~\cite{srivastava2018large,fillion2022gate,koyama2018electric,PRL2019Vedmedenko},
the magnonic $\varphi$ Josephson junction is realizable by tuning coherent coupling appropriately.
Moreover, applying microwave to each FM continuously, the loss of magnons due to dissipation  is precisely balanced by the injection of magnons  achieved through microwave pumping. Therefore,  spins in each FM continue to precess coherently, and the synchronized precession of  the magnonic $\varphi$ Josephson junctions  remains stable.
Thus,  our theoretical prediction is within experimental reach  with current device and measurement techniques
through magnetization measurement.

For the observation of our theoretical predictions,
a few comments are in order. 
It has been established in experiments in which one
attaches a heavy metal (e.g., platinum) to FMs that
magnon currents can be measured electrically 
by the inverse spin Hall effect~\cite{LongMagnon2015NatPhys}.
This is a promising way 
to observe the effects predicted in this work.
Here it should be noted that 
the coherence length of magnons in yttrium iron garnet (YIG) 
reaches  the order of $10\, \mu$m  (i.e., macroscopic)~\cite{MagnonCoherenceLength,ReviewCoherentMagnon2021},
and experimental observation of a magnonic Josephson effect 
in YIG
and that of a magnon supercurrent in YIG
have been reported in Refs.~\cite{ExpMagnonJosephson}  
and~\cite{MagnonSupercurrent}, respectively,
by using Brillouin light scattering spectroscopy.
Hence we expect that YIG 
is one of the best platforms 
for the realization of our proposed setup.
Note that 
the magnitude of the non-local damping can be comparable to 
the local damping in general~\cite{Damping_Heinrich2003},
and recent experimental studies reported that the value of the local damping 
for YIG reaches the order of MHz~\cite{YIGdamping2011}.

\section{Conclusion}
\label{sec:VII}

We have investigated the effect of non-local damping on magnetic junctions and found that  it serves as the key ingredient  for the synchronized precession and gives rise to a magnonic $\varphi$ Josephson junction.
The spacer layer-mediated interaction between the two FMs in the junctions consists of coherent coupling and non-local damping, and it becomes non-Hermitian due to non-local damping. Tuning them appropriately, coherent spin precession in each FM is synchronized  by non-local damping as time advances and forms a $\varphi$ Josephson junction,
where the relative precession angle $\varphi$ decreases monotonically  from $ |\varphi| =\pi$ to  $ |\varphi| =\pi/2$
as the magnitude of the magnon-magnon interaction increases, with both limiting values being entirely universal.
The magnon currents in the  junction  exhibits rectification and gives rise to a magnonic diode effect.
Applying microwaves to each FM continuously, the junction reaches  the nonequilibrium steady state where 
the loss of magnons due to dissipation  is precisely balanced by the injection of magnons  achieved through microwave pumping. Hence, spins in each FM continue to precess coherently, and the synchronized precession of the left and the right magnetization remains stable.

\acknowledgements

This work was supported by the Georg H. Endress Foundation and by the Swiss National Science Foundation, and NCCR SPIN (grant number 51NF40-180604). K.N. acknowledges support by JSPS KAKENHI Grants  No. JP22K03519.





\appendix

\section{Magnonic Josephson equations}
\label{sec:SI}

In this Appendix, we provide details  on the derivation of the magnonic Josephson equations.
The effective non-Hermitian Hamiltonian  $ {\mathcal{H}} $ for the magnonic Josephson junction is given as~\cite{Ji_NHdiode,KKPD,KPD} $ {\mathcal{H}}= {\mathcal{H}}_{\text{L}} +{\mathcal{H}}_{\text{R}}+V+U_{\text{L}}+U_{\text{R}}$ with (see also main text) $ {\mathcal{H}}_{\text{L}}=\hbar(\Omega - i\gamma) a_{\text{L}}^{\dagger}a_{\text{L}}  $ and $ {\mathcal{H}}_{\text{R}} = \hbar(\Omega - i\gamma) a_{\text{R}}^{\dagger}a_{\text{R}} $, where $  a_{\text{L(R)}}^{(\dagger)}  $ represents the magnon annihilation (creation) operator for the zero wavenumber mode (i.e., spatially uniform mode) in the left (right) FM. This provides the time evolution of each operator as
\begin{widetext}
\begin{subequations}
\begin{align}
i \hbar {\dot{a}}_{\text{L}}(t)
=& \hbar(\Omega - i\gamma) a_{\text{L}}(t)
  +({\mathcal{J}}-i\hbar G/2)a_{\text{R}}(t)
  +U a_{\text{L}}^{\dagger}(t) a_{\text{L}}(t)a_{\text{L}}(t), 
\label{eqn:SM1a} \\
i \hbar {\dot{a}}_{\text{R}}(t)
=& \hbar(\Omega - i\gamma) a_{\text{R}}(t)
+({\mathcal{J}}^{*}-i\hbar G/2)a_{\text{L}}(t)
+U a_{\text{R}}^{\dagger}(t) a_{\text{R}}(t)a_{\text{R}}(t).
\label{eqn:SM1b}  
\end{align}
\end{subequations}
Here, assuming a macroscopic coherent magnon state, thereby using the semiclassical approximation, we replace the operators $ a_{\text{L}}(t) $ and $ a_{\text{R}}(t) $ by their expectation values as
$ \langle a_{\text{L}}(t) \rangle 
=\sqrt{N_{\text{L}}(t)} 
{\text{e}}^{i\theta_{\text{L}}(t)}  $
and
$ \langle a_{\text{R}}(t) \rangle
=\sqrt{N_{\text{R}}(t)} 
{\text{e}}^{i\theta_{\text{R}}(t)}  $,
respectively, where $ N_{\text{L(R)}}(t) \in {\mathbb{R}} $ represents the number of coherent magnons
for each site in the left (right) FM and  $ \theta_{\text{L(R)}}(t) \in {\mathbb{R}} $ is the phase. Defining the relative phase as
\begin{equation}
\theta(t)= \theta_{\text{R}}(t)-\theta_{\text{L}}(t),
\end{equation}
Eqs.~\eqref{eqn:SM1a} and~\eqref{eqn:SM1b}  for $ N_{\text{L(R)}}(t)\neq0 $ become
\begin{subequations}
\begin{align}
i \hbar \Big( \frac{1}{2} \frac{{\dot{N}}_{\text{L}}}{N_{\text{L}}} +i {\dot{\theta}}_{\text{L}} \Big)
&=\hbar(\Omega - i\gamma) +UN_{\text{L}} +\Big[
{\mathcal{J}}_{\text{Re}} +i \Big( {\mathcal{J}}_{\text{Im}}-\frac{\hbar}{2}G \Big) \Big] \sqrt{\frac{N_{\text{R}}}{N_{\text{L}}}}
{\text{e}}^{i \theta},
\label{eqn:SM2a}  \\
i \hbar \Big( \frac{1}{2} \frac{{\dot{N}}_{\text{R}}}{N_{\text{R}}}
+i {\dot{\theta}}_{\text{R}} \Big) &=\hbar(\Omega - i\gamma)
+UN_{\text{R}} +\Big[{\mathcal{J}}_{\text{Re}}-i \Big( {\mathcal{J}}_{\text{Im}}+\frac{\hbar}{2}G
\Big) \Big] \sqrt{\frac{N_{\text{L}}}{N_{\text{R}}}}{\text{e}}^{-i \theta},
\label{eqn:SM2b}  
\end{align}
\end{subequations}
 where  ${\text{Re}}({\mathcal{J}})= {\mathcal{J}}_{\text{Re}} \in {\mathbb{R}} $  and ${\text{Im}}({\mathcal{J}})= {\mathcal{J}}_{\text{Im}} \in {\mathbb{R}} $. Real and imaginary parts of  Eqs.~(\ref{eqn:SM2a}) and~(\ref{eqn:SM2b})
provide
\begin{subequations}
\begin{align}
- \hbar {\dot{\theta}}_{\text{L}} &=\hbar \Omega  +UN_{\text{L}} +\Big[ {\mathcal{J}}_{\text{Re}}  \cos \theta - \Big(
{\mathcal{J}}_{\text{Im}}-\frac{\hbar}{2}G \Big) \sin\theta \Big] \sqrt{\frac{N_{\text{R}}}{N_{\text{L}}}},
\label{eqn:SM3a}  \\
\frac{\hbar}{2} \frac{{\dot{N}}_{\text{L}}}{N_{\text{L}}} &=-\hbar \gamma +\Big[
{\mathcal{J}}_{\text{Re}} \sin \theta + \Big( {\mathcal{J}}_{\text{Im}}-\frac{\hbar}{2}G
\Big) \cos \theta \Big] \sqrt{\frac{N_{\text{R}}}{N_{\text{L}}}},
\label{eqn:SM3b}  \\
- \hbar {\dot{\theta}}_{\text{R}} &=\hbar \Omega  +UN_{\text{R}} +\Big[ {\mathcal{J}}_{\text{Re}} {{\cos}}\theta - \Big(
{\mathcal{J}}_{\text{Im}}+\frac{\hbar}{2}G \Big)\sin\theta\Big] \sqrt{\frac{N_{\text{L}}}{N_{\text{R}}}},
\label{eqn:SM3c}  \\
\frac{\hbar}{2} \frac{{\dot{N}}_{\text{R}}}{N_{\text{R}}} &=-\hbar \gamma
-\Big[ {\mathcal{J}}_{\text{Re}} \sin \theta + \Big( {\mathcal{J}}_{\text{Im}}+\frac{\hbar}{2}G
\Big) \cos \theta \Big] \sqrt{\frac{N_{\text{L}}}{N_{\text{R}}}},
\label{eqn:SM3d} 
\end{align}
\end{subequations}
where Eq.~(\ref{eqn:SM3a}) is the real part of Eq.~(\ref{eqn:SM2a}),
Eq.~(\ref{eqn:SM3b}) is the imaginary part of Eq.~(\ref{eqn:SM2a}), Eq.~(\ref{eqn:SM3c}) is the real part of Eq.~(\ref{eqn:SM2b}), and Eq.~(\ref{eqn:SM3d}) is the imaginary part of Eq.~(\ref{eqn:SM2b}).
Taking the difference between Eqs.~(\ref{eqn:SM3a}) and~(\ref{eqn:SM3c}), those are rewritten as
\begin{subequations}
\begin{align}
  {\dot{\theta}}(t) =&  \frac{{\mathcal{J}}_{\text{Re}}}{\hbar} \cos \theta \Big(
\sqrt{\frac{N_{\text{R}}}{N_{\text{L}}}} -\sqrt{\frac{N_{\text{L}}}{N_{\text{R}}}}
\Big) + \sin \theta \Big[ \Big(\frac{{\mathcal{J}}_{\text{Im}}}{\hbar} +\frac{G}{2}\Big)
\sqrt{\frac{N_{\text{L}}}{N_{\text{R}}}} -\Big(\frac{{\mathcal{J}}_{\text{Im}}}{\hbar}
-\frac{G}{2}\Big)  \sqrt{\frac{N_{\text{R}}}{N_{\text{L}}}} \Big] +u(N_{\text{L}}-N_{\text{R}}), \label{eqn:SM4a}  \\
{\dot{N}}_{\text{L}}(t) =& -2 \gamma N_{\text{L}} +2\Big[ \frac{{\mathcal{J}}_{\text{Re}}}{\hbar}
\sin \theta +\Big(\frac{{\mathcal{J}}_{\text{Im}}}{\hbar} -\frac{G}{2}\Big)  \cos \theta
\Big] \sqrt{N_{\text{L}}N_{\text{R}}},\label{eqn:SM4b}  \\
{\dot{N}}_{\text{R}}(t) =& -2 \gamma  N_{\text{R}} -2\Big[ \frac{{\mathcal{J}}_{\text{Re}}}{\hbar} \sin\theta
+\Big(\frac{{\mathcal{J}}_{\text{Im}}}{\hbar} +\frac{G}{2}\Big)  \cos \theta \Big] \sqrt{N_{\text{L}}N_{\text{R}}}.
\label{eqn:SM4c} 
\end{align}
\end{subequations}
In this study, we assume to continuously apply microwaves to each FM. The coherent magnon state,
the key ingredient for the magnonic Josephson effect, is realized by microwave pumping. Under microwave pumping, coherent magnons are injected into each FM at the rate of $\mathcal{P}$~\cite{KPD}. Taking this effect of the magnon injection  through microwave pumping into account, Eqs.~\eqref{eqn:SM4b} and~\eqref{eqn:SM4c} become
\begin{subequations}
\begin{align}
{\dot{N}}_{\text{L}}(t) =& -2 \gamma N_{\text{L}} +2\Big[ \frac{{\mathcal{J}}_{\text{Re}}}{\hbar} \sin \theta
+\Big(\frac{{\mathcal{J}}_{\text{Im}}}{\hbar} -\frac{G}{2}\Big)  \cos \theta \Big]
\sqrt{N_{\text{L}}N_{\text{R}}}+\mathcal{P}, \label{eqn:SM4b2}  \\
{\dot{N}}_{\text{R}}(t) =& -2 \gamma  N_{\text{R}} -2\Big[\frac{{\mathcal{J}}_{\text{Re}}}{\hbar}\sin \theta
+\Big(\frac{{\mathcal{J}}_{\text{Im}}}{\hbar}+\frac{G}{2}\Big)  \cos \theta \Big] \sqrt{N_{\text{L}}N_{\text{R}}}+\mathcal{P}.
\label{eqn:SM4c2} 
\end{align}
\end{subequations}
Finally, the magnonic Josephson equations under microwave pumping in the presence of non-local damping
are summarized as
\begin{subequations}
\begin{align}
{\dot{\theta}}(t)=& \frac{{\mathcal{J}}_{\text{Re}}}{\hbar} \cos \theta(t) \Big( \sqrt{\frac{N_{\text{R}}(t)}{N_{\text{L}}(t)}}
-\sqrt{\frac{N_{\text{L}}(t)}{N_{\text{R}}(t)}} \Big)
+ \sin \theta(t) \Big[ \Big(\frac{{\mathcal{J}}_{\text{Im}}}{\hbar} +\frac{G}{2}\Big) \sqrt{\frac{N_{\text{L}}(t)}{N_{\text{R}}(t)}}
-\Big(\frac{{\mathcal{J}}_{\text{Im}}}{\hbar} -\frac{G}{2}\Big)  \sqrt{\frac{N_{\text{R}}(t)}{N_{\text{L}}(t)}}
\Big] +u[N_{\text{L}}(t)-N_{\text{R}}(t)], \label{eqn:SM4a2}  \\
{\dot{N}}_{\text{L}}(t) =& -2 \gamma N_{\text{L}}(t) +2\Big[ \frac{{\mathcal{J}}_{\text{Re}}}{\hbar}
\sin \theta(t) +\Big(\frac{{\mathcal{J}}_{\text{Im}}}{\hbar} -\frac{G}{2}\Big)  \cos \theta(t) \Big] \sqrt{N_{\text{L}}(t)N_{\text{R}}(t)} +{\mathcal{P}}, \label{eqn:SM4b2}  \\
{\dot{N}}_{\text{R}}(t) =& -2 \gamma  N_{\text{R}}(t) -2\Big[\frac{{\mathcal{J}}_{\text{Re}}}{\hbar} \sin \theta(t)
+\Big(\frac{{\mathcal{J}}_{\text{Im}}}{\hbar} +\frac{G}{2}\Big) \cos \theta(t)\Big] \sqrt{N_{\text{L}}(t)N_{\text{R}}(t)}
+{\mathcal{P}}. \label{eqn:SM4c2} 
\end{align}
\end{subequations}
\end{widetext}

\section{Nonequilibrium steady state for finite magnon interaction}
\label{sec:SIV}

In this Appendix, we provide details  on the derivation of the equation for $ \varphi$ as a function of the magnon-magnon interaction. For \begin{equation}
{\mathcal{J}}_{\text{Re}}  =0 \  \ \text{and} \  \  {\mathcal{J}}_{\text{Im}}/\hbar = G/2,
\label{eqn:SM9}
\end{equation}
Eqs.~\eqref{eqn:SM4a2}-\eqref{eqn:SM4c2} become
\begin{subequations}
\begin{align}
  {\dot{\theta}}(t)=&  G  \sqrt{\frac{N_{\text{L}}(t)}{N_{\text{R}}(t)}}
\sin \theta(t) +u[N_{\text{L}}(t)-N_{\text{R}}(t)], \label{eqn:SM10a}  \\
{\dot{N}}_{\text{L}}(t) =& -2 \gamma N_{\text{L}}(t) +{\mathcal{P}}, \label{eqn:SM10b}  \\
{\dot{N}}_{\text{R}}(t) =& -2 \gamma  N_{\text{R}}(t)-2G \sqrt{N_{\text{L}}(t)N_{\text{R}}(t)} \cos \theta(t)
+{\mathcal{P}}.
\label{eqn:SM10c} 
\end{align}
\end{subequations}
Under microwave pumping $\mathcal{P}$,
the nonequilibrium steady state  $ {\dot{\theta}}(t) ={\dot{N}}_{\text{L}}(t) = {\dot{N}}_{\text{R}}(t) =0  $
is realized, where $\theta(t) $, $ N_{\text{L}}(t)$, and $ N_{\text{R}}(t) $ approach asymptotically to time-independent constant as time advances, $\varphi=\lim_{t \to \infty}  \theta(t)$ and $N_{\text{L,R}}^{\infty} =\lim_{t \to \infty} N_{\text{L,R}}(t)$. Thus, from Eqs.~\eqref{eqn:SM10a}, ~\eqref{eqn:SM10b} and~\eqref{eqn:SM10c}, for $ u \neq 0$, we get 
\begin{subequations}
\begin{align}
&N_{\text{L}}^{\infty}-N_{\text{R}}^{\infty}=-\frac{G}{u}
\sqrt{\frac{N_{\text{L}}^{\infty} }{ N_{\text{R}}^{\infty}}}
\sin \varphi, \label{eqn:SMsin}\\
&N_{\text{L}}^{\infty}  =\frac{\mathcal{P}}{2 \gamma},  
\label{eqn:SMNLinf} \\
&N_{\text{L}}^{\infty}-N_{\text{R}}^{\infty} =\frac{G}{\gamma} \sqrt{N_{\text{L}}^{\infty} N_{\text{R}}^{\infty}} \cos \varphi,
\label{eqn:SMcos} 
\end{align}
\end{subequations}
From Eq.~\eqref{eqn:SMcos}, we get the ratio between the number of coherent magnons in the left and right FMs,
\begin{align}
\sqrt{\frac{N_{\text{R}}^{\infty}}{N_{\text{L}}^{\infty}}} =  \sqrt{1+ \left(\frac{G}{2\gamma}\cos \varphi\right)^2} -  \frac{G}{2\gamma}\cos \varphi.
\end{align}
Combining Eq.~\eqref{eqn:SMsin}  and Eq.~\eqref{eqn:SMcos}, we obtain
\begin{align}
& \tan \varphi =-\frac{u}{\gamma} N_{\text{R}}^{\infty}, \label{eqn:SMtan} \\
&(N_{\text{L}}^{\infty}-N_{\text{R}}^{\infty})^2 =-\frac{G^2}{\gamma u} N_{\text{L}}^{\infty} \frac{\sin(2\varphi)}{2}
\label{eqn:SMsin2}.
\end{align}
If $u>0$, the solution exists only for $\tan\varphi<0$. Substituting $N_{\text{R}}^{\infty}$ from Eq. ~\eqref{eqn:SMtan} into Eq.~\eqref{eqn:SMsin2}, we arrive at the implicit equation for $\tan \varphi$:
\begin{equation}
\Big( \frac{\gamma}{u} \tan \varphi +N_{\text{L}}^{\infty}\Big)^2 =- \frac{G^2}{\gamma u} N_{\text{L}}^{\infty} \frac{ \tan \varphi}{1+ \tan^2\varphi}.
\label{eqn:SMtantan} 
\end{equation}
This equation can be brought into the form of a quartic equation displayed in the main text: 
\begin{equation}
\tan^4\varphi +2 \tilde{u} \tan^3\varphi 
+(1+ {\tilde{u}}^2) \tan^2\varphi    
+\left[2+\left(\frac{G}{\gamma}\right)^2\right]
 {\tilde{u}} \tan \varphi + {\tilde{u}}^2=0, 
 \label{eqn:SMvarphiEq} 
\end{equation}
where we introduced the rescaled dimensionless magnon-magnon interaction parameter
$  {\tilde{u}}=(u/\gamma) N_{\text{L}}^{\infty}$. We note that the solution is possible only for $\tan \varphi <0$ ($\tan \varphi>0$) for $u>0$ ($u<0$). We note that for a quartic polynomial one can find an explicit solution. However, it is too involved to be displayed here. In the main text we give the analytical expressions for $\varphi$ for the limiting cases of small and large ${\tilde{u}}$.

\section{Dynamics with different pumping rates}
\label{sec:SII}
In this Appendix, we consider the case where $\mathcal{J}_{\text{Re}} = 0$, $\mathcal{J}_{\text{Im}}/\hbar = G/2$, and $u = 0$, with different pumping rates for two FMs. By solving the dynamics analytically, we show that our results do not rely on the initial conditions.

When we only pump the left FM with rate $\mathcal{P}$,
the coupled dynamics is described by the following equations: 
  \(  \al{  \dot{\theta}&=G\sqrt{\frac{N_{\text{L}}}{N_{\text{R}}}} \sin\theta, \\
                    \dot{N}_{\text{L}}&=-2\gamma N_{\text{L}} +\mathcal{P}, \\
                    \dot{N}_{\text{R}}&=-2\gamma N_{\text{R}} - 2G\sqrt{N_{\text{L}}N_{\text{R}}} \cos\theta. }    \)
We now solve these equations analytically. We first introduce $n_{{\text{L}}/{\text{R}}}(t)=e^{2\gamma t} N_{{\text{L}}/{\text{R}}}(t)$, which allows us to rewrite the equations into the following form: 
\(   \dot{n}_{\text{L}}= e^{2\gamma t} \mathcal{P} , \;\; \dot{\theta}=G\sqrt{\frac{n_{\text{L}}}{n_{\text{R}}}} \sin\theta,\;\; \dot{n}_{\text{R}}=-2G\sqrt{n_{\text{R}}n_{\text{L}}} \cos\theta.    \)
The first equation can be solved as
\(  n_{\text{L}}(t)= N_{\text{L}}(0) + \frac{ \mathcal{P}  }{2\gamma}(e^{2\gamma t}-1).   \)
The last two equations leads to 
\(  \dv{\theta}{n_{\text{R}}}=-\frac{1}{2} \frac{\tan\theta}{n_{\text{R}}} \)
and hence
 \( |\sin\theta(t)| \sqrt{n_{\text{R}}(t)}=|\sin\theta(0)| \sqrt{N_{\text{R}}(0)}.    \)
 We then obtain the equation for $\theta(t)$
 under the initial condition $  \theta(0) \neq 0$  and $  \theta(0) \neq  \pm \pi$ as
            \(  \dv{\theta}{t}= \frac{G\sqrt{n_{\text{L}}(t)}}{|\sin\theta(0)|\sqrt{N_{\text{R}}(0)}} \sin\theta |\sin\theta|.   \)
           We remark that the above equation can be solved analytically because the integral of $\sqrt{n_{\text{L}}(t)}$ can be performed analytically. 
           Henceforth,  for our purpose, we focus on the case where $t\gg 1/(2\gamma)$ and hence $n_{\text{L}}(t) \approx (\mathcal{P}/2\gamma)e^{2\gamma t}$.

Here, without loss of generality, we consider the case of $\sin\theta(0)>0$ and look for the solution that satisfies $\sin\theta(t)>0$: One can similarly solve for the case when $\sin\theta(0)<0$.  
 We then get
       \( \tan\theta(t)= \frac{1}{\cot\theta(0)- \alpha e^{\gamma t} } \)
 and hence    
      \( \theta(t)=\arctan \left[ \frac{1}{\cot\theta(0)- \alpha e^{\gamma t} } \right],  \)
       where $\alpha$ is the constant defined by
       \(  \alpha\equiv \frac{(G/\gamma)\sqrt{\mathcal{P}/(2\gamma)}}{\sin\theta(0) \sqrt{N_{\text{R}}(0)}} .  \)
In the large $t\rightarrow \infty$ limit, we can drop $\cot\theta(0)$ in the expression. Then the expression of $\theta(t)$ is reduced to 
\( {  \theta(t)= \pi - \frac{1}{\alpha} e^{-\gamma t} .  }   \)
One can also easily write down the expression of $n_{\text{R}}(t)$ as
\(  { \sqrt{n_{\text{R}}(t)}= \alpha |\sin\theta(0)| \sqrt{N_{\text{R}}(0)} e^{2\gamma t}.  }  \)
 At large $t\rightarrow \infty$, we get
\( { \sqrt{\frac{N_{\text{R}}^{\infty}}{N_{\text{L}}^{\infty}}}=\frac{G}{\gamma}.}  \)
It should be noted that this final value is independent of the initial condition and also the pumping rate $\mathcal{P}$.

When we pump the two FMs at the same time, the physics is very similar to the case that we discussed above. The only difference is that $N_{\text{R}}^{\infty}$ is different. The equations are given by
 \(  \al{  \dot{\theta}&=G\sqrt{\frac{N_{\text{L}}}{N_{\text{R}}}} \sin\theta, \\
                    \dot{N}_{\text{L}}&=-2\gamma N_{\text{L}} +\mathcal{P}_{\text{L}}, \\
                    \dot{N}_{\text{R}}&=-2\gamma N_{\text{R}} - 2G\sqrt{N_{\text{L}}N_{\text{R}}} \cos\theta+\mathcal{P}_{\text{R}}, }    \)
  where $\mathcal{P}_{\text{L}}$ and $\mathcal{P}_{\text{R}}$ are the pumping rates of the two FMs. 
  Introducing the ratio
                    \( \beta\equiv \sqrt{\frac{N_{\text{R}}^{\infty}}{N_{\text{L}}^{\infty}}}, \)
    it is determined by the equation
 \( \beta^2- \frac{G}{\gamma} \beta - \frac{\mathcal{P}_{\text{R}}}{\mathcal{P}_{\text{L}}}=0 \)
 and hence
                   \( \beta= \frac{G/\gamma+\sqrt{(G/\gamma)^2+4\mathcal{P}_{\text{R}}/\mathcal{P}_{\text{L}}} }{2}  \)
                    at $t\rightarrow \infty$. 
 Here, without loss of generality, we consider the case of $\sin\theta(0)>0$ and look for the solution that satisfies $\sin\theta(t)>0$. 
                    Similar to our previous conclusion, $\theta(t)$ also approaches $\pi$ exponentially fast but now with a different exponent as
                    \(  { \pi-\theta(t)\propto e^{- (G/\beta)t}.   }  \)
                We note that in the absence of the pumping of the right FM $\mathcal{P}_{\text{R}}=0$, we get $\beta=G/\gamma$ and $ \pi-\theta(t)\propto e^{- \gamma t}$. This agrees with what we obtained before.

\begin{figure}[!t]
\begin{center}
\includegraphics[width=8.8cm,clip]{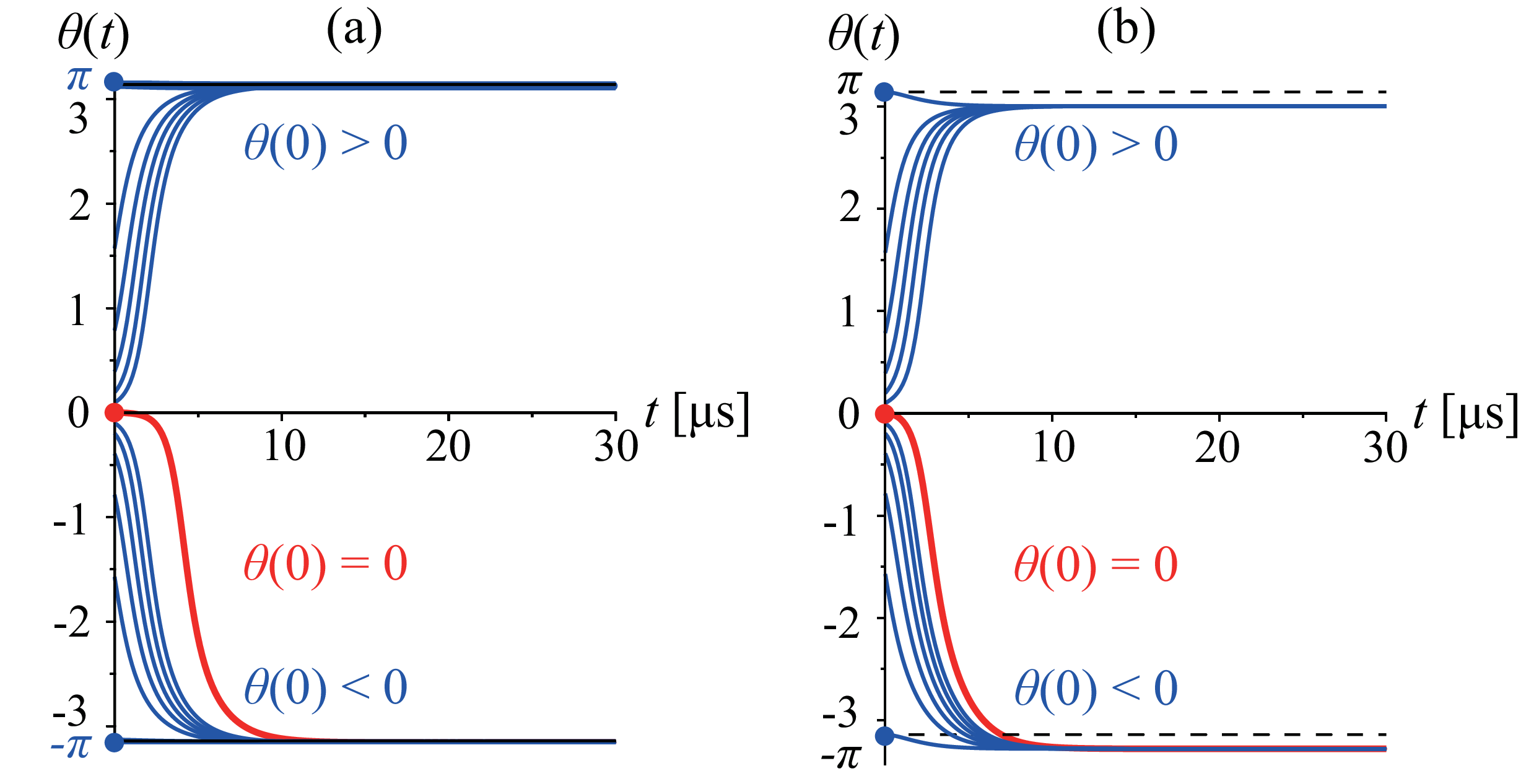}
\caption{
Plots of  the relative phase $\theta(t)$ as a function of time in the absence of the magnon-magnon interaction, 
i.e., $u=0$, for the initial condition $ \theta(0)=0 $ and $ |\theta(0)|= \pi/2^n $ with $n=0,1,...,5$
obtained by numerically solving  Eqs.~\eqref{eqn:SM4a2}-\eqref{eqn:SM4c2}. The parameter values are the same as in Fig.~2 of the main text, e.g., $G=1$ MHz, except for (a) $C_1=C_2=10$ kHz and (b) $C_1=100$ kHz and $C_2=10$ kHz. Even in the presence of such perturbation $C_{1(2)}$,  the synchronized precession of  the left and the right magnetization,  $ {\dot{\theta}}(t) =0$, remains valid,
and
the magnitude of $\varphi$ is still bounded as 
Eq.~\eqref{eqn:SMboundValue} 
where the value of $ \varphi $ slightly changes depending on the magnitude of the deviation ($C_1$ and $C_2$).
Under the initial condition $\theta(0)=0$, 
we get $ \varphi \neq 0$ for $ C_1 \neq 0$.
}
\label{fig:CD}
\end{center}
\end{figure}

\begin{figure}
\begin{center}
\includegraphics[width=8.8cm,clip]{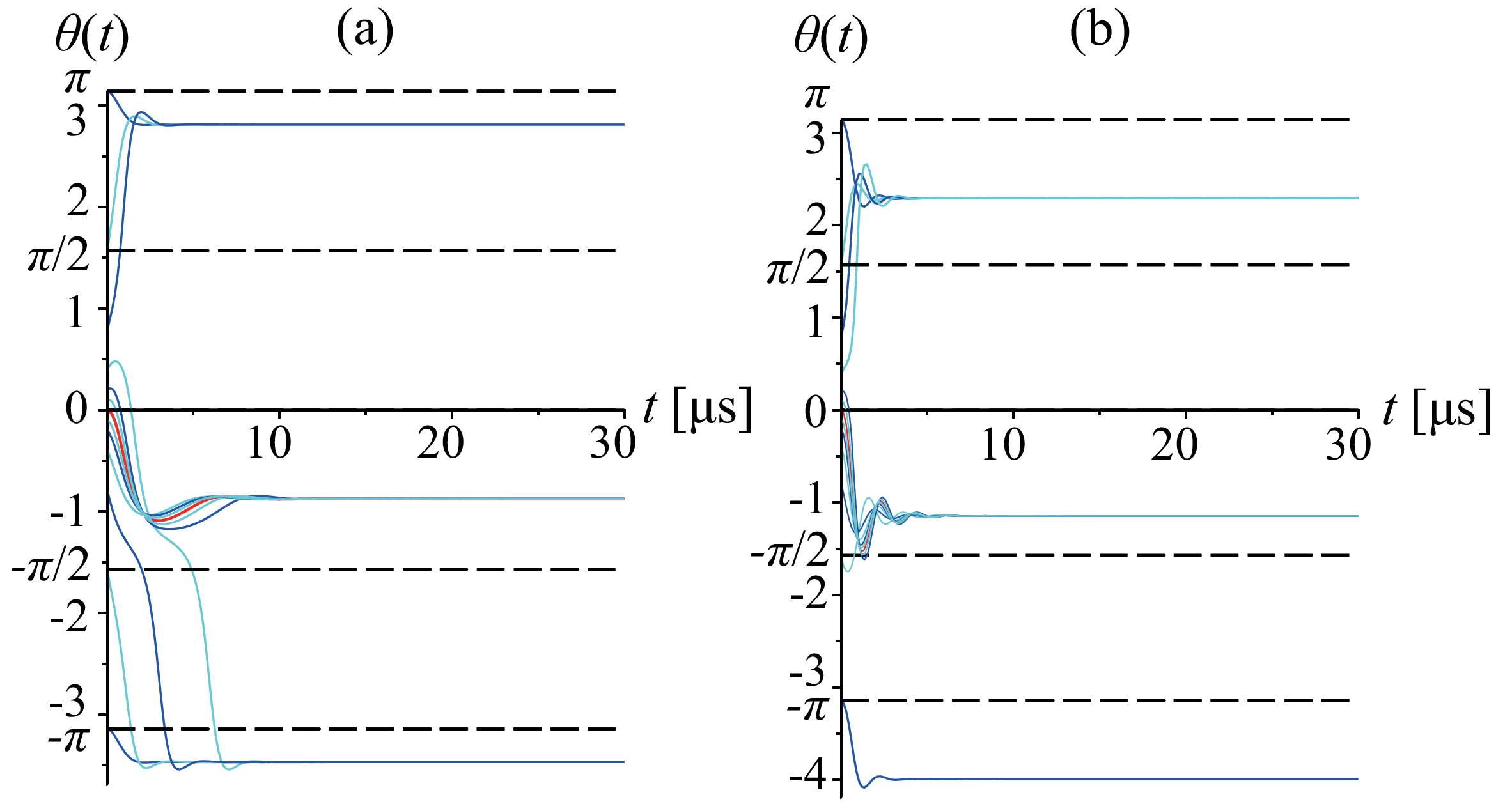}
\caption{
Plots of the relative phase $\theta(t)$ as a function of time
obtained by numerically solving  Eqs.~\eqref{eqn:SM4a2}-\eqref{eqn:SM4c2}
for the same parameter values as in 
Fig.~\ref{fig:CD}
except for 
(a) $C_1=1$ MHz and $C_2=10$ kHz and 
(b) $C_1=C_2=1$ MHz.
The magnitude of $\varphi$ is no longer bounded as 
Eq.~\eqref{eqn:SMboundValue},
and the value of $\varphi$ changes
depending on the initial condition $\theta(0)$.
Still,
the synchronized precession of the left and the right magnetization,  
$ {\dot{\theta}}(t) =0$, 
remains valid.
Under the initial condition $\theta(0)=0$, 
we get $ \varphi \neq 0$
for $ C_1 \neq 0$.
}
\label{fig:6466}
\end{center}
\end{figure}

\section{Robustness of synchronized precession}
\label{sec:SIII}

In the main text, to seek for the junction that exhibits a rectification effect characterized by $  I_{\text{L}}(t) =0 $,
we consider the case 
\begin{equation} 
{\mathcal{J}}_{\text{Re}}  =0
\  \ \text{and} \   \  {\mathcal{J}}_{\text{Im}}/\hbar = G/2 >0
\label{eqn:9SM}
\end{equation}
and find that 
the magnitude of $\varphi$ is bounded as 
\begin{equation} 
{\pi}/{2}<| \varphi | \leq \pi.
\label{eqn:SMboundValue}
\end{equation}

\begin{figure}
[!b]
\begin{center}
\includegraphics[width=4.9cm,clip]{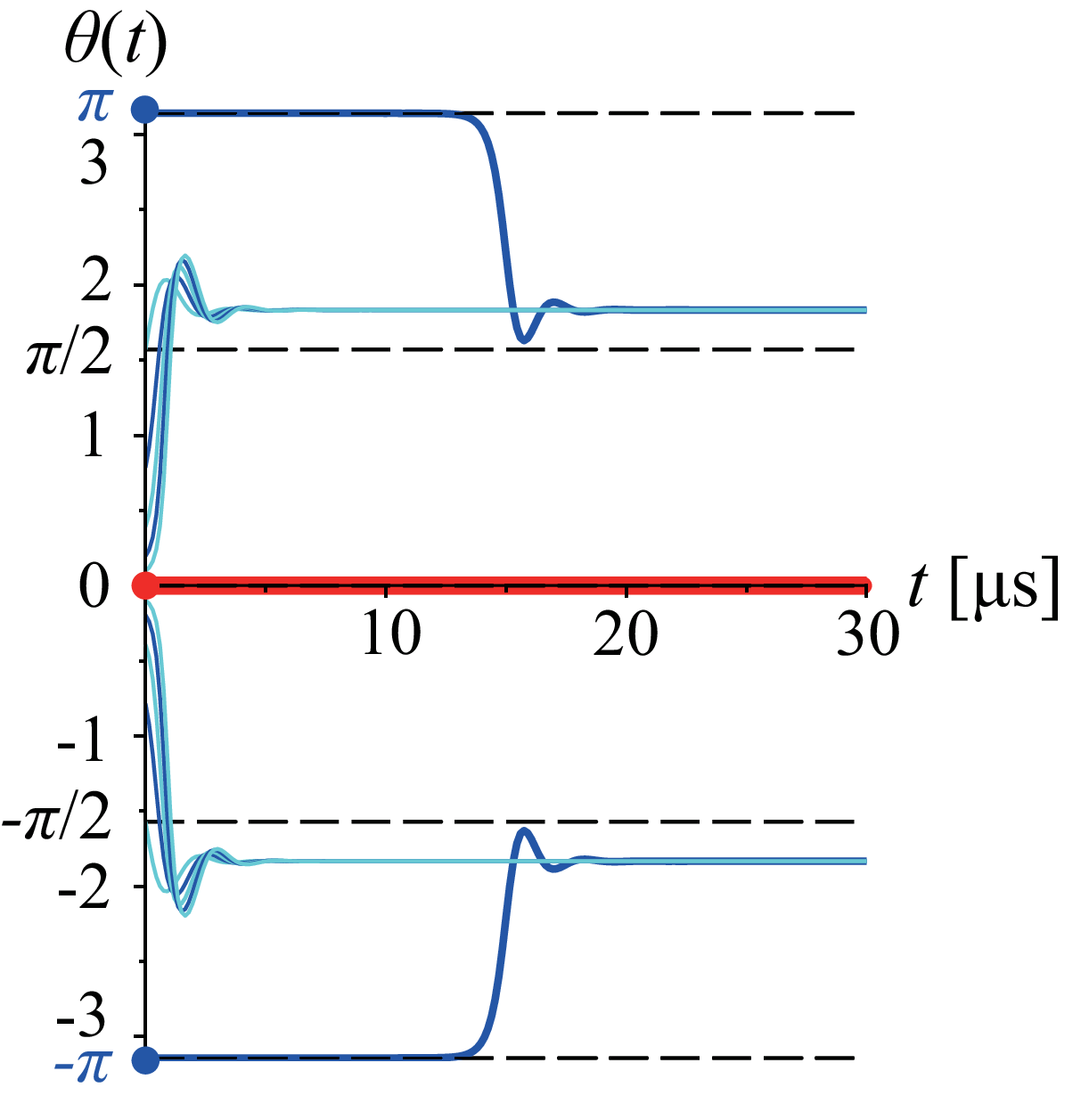}
\caption{
Plots of the relative phase $\theta(t)$ as a function of time
obtained by numerically solving  Eqs.~\eqref{eqn:SM4a2}-\eqref{eqn:SM4c2}
for the same parameter values as in 
Fig.~\ref{fig:6466}
except for 
$C_1=0$ and $C_2=1$ MHz.
The magnitude of $\varphi$ is bounded as 
Eq.~\eqref{eqn:SMboundValue},
and
the synchronized precession of the left and the right magnetization,  $ {\dot{\theta}}(t) =0$, 
remains valid.
Under the initial condition $\theta(0)=0$, 
we get $ \varphi = 0$ for $ C_1 = 0$. 
}
\label{fig:06}
\end{center}
\end{figure}

In this Appendix, although  the rectification effect ceases to work  and it becomes  $  I_{\text{L}}(t) \neq0 $, we numerically show that the synchronized precession of the left and the right magnetization,  $ {\dot{\theta}}(t) =0$, 
remains valid even if the parameter values deviate  from Eq.~\eqref{eqn:9SM}. 
For this, we consider cases with
\begin{equation}
{\mathcal{J}}_{\text{Re}}/\hbar  = C_1 \  \ \text{and} \  \ {\mathcal{J}}_{\text{Im}}/\hbar = G/2+C_2,
\label{eqn:9SM2}
\end{equation}
where $ C_{1(2)} $ is the constant that satisfies the condition
$ |\mathcal{J}| \ll \hbar \Omega $
for weakly spin-exchange coupled junctions.
For  $ | C_{1}| \ll G$,
Fig.~\ref{fig:CD} shows that 
the magnitude of $\varphi$ is bounded as 
Eq.~\eqref{eqn:SMboundValue}.
For  $ | C_{1}| \sim G$, i.e.,
when the value of $C_1$ increases and amounts to of the order of $G$,
Fig.~\ref{fig:6466} shows that
the magnitude of $\varphi$ is no longer bounded as 
Eq.~\eqref{eqn:SMboundValue}.
For  $ | C_{2}| \sim G$ and $C_1=0 $,
Fig.~\ref{fig:06} shows that
the value of $\varphi$ differs from $\pm \pi$,
but still its magnitude is bounded as 
Eq.~\eqref{eqn:SMboundValue}.
We emphasize that 
the synchronized precession of the left and the right magnetization,  
$ {\dot{\theta}}(t) =0$, 
remains valid
in all of these cases (Figs.~\ref{fig:CD}-\ref{fig:06}).
We also note that 
under the initial condition $\theta(0)=0$,
Figs.~\ref{fig:CD}-\ref{fig:06} with Fig.~2 of the main text
show that we get $ \varphi \neq 0$ for $ C_1 \neq 0$, 
whereas $ \varphi = 0$ for $ C_1 = 0$. 
Also in this sense, the point $\theta(0)=0$ is instable.

\bibliography{PumpingRef}

\begin{thebibliography}{67}%
\makeatletter
\providecommand \@ifxundefined [1]{%
 \@ifx{#1\undefined}
}%
\providecommand \@ifnum [1]{%
 \ifnum #1\expandafter \@firstoftwo
 \else \expandafter \@secondoftwo
 \fi
}%
\providecommand \@ifx [1]{%
 \ifx #1\expandafter \@firstoftwo
 \else \expandafter \@secondoftwo
 \fi
}%
\providecommand \natexlab [1]{#1}%
\providecommand \enquote  [1]{``#1''}%
\providecommand \bibnamefont  [1]{#1}%
\providecommand \bibfnamefont [1]{#1}%
\providecommand \citenamefont [1]{#1}%
\providecommand \href@noop [0]{\@secondoftwo}%
\providecommand \href [0]{\begingroup \@sanitize@url \@href}%
\providecommand \@href[1]{\@@startlink{#1}\@@href}%
\providecommand \@@href[1]{\endgroup#1\@@endlink}%
\providecommand \@sanitize@url [0]{\catcode `\\12\catcode `\$12\catcode `\&12\catcode `\#12\catcode `\^12\catcode `\_12\catcode `\%12\relax}%
\providecommand \@@startlink[1]{}%
\providecommand \@@endlink[0]{}%
\providecommand \url  [0]{\begingroup\@sanitize@url \@url }%
\providecommand \@url [1]{\endgroup\@href {#1}{\urlprefix }}%
\providecommand \urlprefix  [0]{URL }%
\providecommand \Eprint [0]{\href }%
\providecommand \doibase [0]{http://dx.doi.org/}%
\providecommand \selectlanguage [0]{\@gobble}%
\providecommand \bibinfo  [0]{\@secondoftwo}%
\providecommand \bibfield  [0]{\@secondoftwo}%
\providecommand \translation [1]{[#1]}%
\providecommand \BibitemOpen [0]{}%
\providecommand \bibitemStop [0]{}%
\providecommand \bibitemNoStop [0]{.\EOS\space}%
\providecommand \EOS [0]{\spacefactor3000\relax}%
\providecommand \BibitemShut  [1]{\csname bibitem#1\endcsname}%
\let\auto@bib@innerbib\@empty
\bibitem [{\citenamefont {Ashida}\ \emph {et~al.}(2020)\citenamefont {Ashida}, \citenamefont {Gong},\ and\ \citenamefont {Ueda}}]{Review_NH_Ashida}%
  \BibitemOpen
  \bibfield  {author} {\bibinfo {author} {\bibfnamefont {Y.}~\bibnamefont {Ashida}}, \bibinfo {author} {\bibfnamefont {Z.}~\bibnamefont {Gong}}, \ and\ \bibinfo {author} {\bibfnamefont {M.}~\bibnamefont {Ueda}},\ }\bibfield  {title} {\enquote {\bibinfo {title} {Non-{H}ermitian physics},}\ }\href {https://doi.org/10.1080/00018732.2021.1876991} {\bibfield  {journal} {\bibinfo  {journal} {Adv. Phys.}\ }\textbf {\bibinfo {volume} {69}},\ \bibinfo {pages} {249} (\bibinfo {year} {2020})},\ \Eprint {http://arxiv.org/abs/arXiv:2006.01837} {arXiv:2006.01837} \BibitemShut {NoStop}%
\bibitem [{\citenamefont {Chumak}\ \emph {et~al.}(2015)\citenamefont {Chumak}, \citenamefont {Vasyuchka}, \citenamefont {Serga},\ and\ \citenamefont {Hillebrands}}]{MagnonSpintronics}%
  \BibitemOpen
  \bibfield  {author} {\bibinfo {author} {\bibfnamefont {A.~V.}\ \bibnamefont {Chumak}}, \bibinfo {author} {\bibfnamefont {V.~I.}\ \bibnamefont {Vasyuchka}}, \bibinfo {author} {\bibfnamefont {A.~A.}\ \bibnamefont {Serga}}, \ and\ \bibinfo {author} {\bibfnamefont {B.}~\bibnamefont {Hillebrands}},\ }\bibfield  {title} {\enquote {\bibinfo {title} {Magnon spintronics},}\ }\href {https://doi.org/10.1038/nphys3347} {\bibfield  {journal} {\bibinfo  {journal} {Nat. Phys.}\ }\textbf {\bibinfo {volume} {11}},\ \bibinfo {pages} {453} (\bibinfo {year} {2015})}\BibitemShut {NoStop}%
\bibitem [{\citenamefont {Chumak}\ \emph {et~al.}(2022)\citenamefont {Chumak}, \citenamefont {Kabos}, \citenamefont {Wu}, \citenamefont {Abert}, \citenamefont {Adelmann}, \citenamefont {Adeyeye}, \citenamefont {{\AA}kerman}, \citenamefont {Aliev}, \citenamefont {Anane}, \citenamefont {Awad} \emph {et~al.}}]{Chumak_Roadmap_SWcomputing}%
  \BibitemOpen
  \bibfield  {author} {\bibinfo {author} {\bibfnamefont {A.~V.}\ \bibnamefont {Chumak}}, \bibinfo {author} {\bibfnamefont {P.}~\bibnamefont {Kabos}}, \bibinfo {author} {\bibfnamefont {M.}~\bibnamefont {Wu}}, \bibinfo {author} {\bibfnamefont {C.}~\bibnamefont {Abert}}, \bibinfo {author} {\bibfnamefont {C.}~\bibnamefont {Adelmann}}, \bibinfo {author} {\bibfnamefont {A.}~\bibnamefont {Adeyeye}}, \bibinfo {author} {\bibfnamefont {J.}~\bibnamefont {{\AA}kerman}}, \bibinfo {author} {\bibfnamefont {F.~G.}\ \bibnamefont {Aliev}}, \bibinfo {author} {\bibfnamefont {A.}~\bibnamefont {Anane}}, \bibinfo {author} {\bibfnamefont {A.}~\bibnamefont {Awad}},  \emph {et~al.},\ }\bibfield  {title} {\enquote {\bibinfo {title} {Advances in magnetics roadmap on spin-wave computing},}\ }\href {https://doi.org/10.1109/TMAG.2022.3149664} {\bibfield  {journal} {\bibinfo  {journal} {IEEE Trans. Magn.}\ }\textbf {\bibinfo {volume} {58}},\ \bibinfo {pages} {1} (\bibinfo {year} {2022})},\ \Eprint {http://arxiv.org/abs/arXiv:2111.00365} {arXiv:2111.00365} \BibitemShut {NoStop}%
\bibitem [{\citenamefont {Yuan}\ \emph {et~al.}(2022)\citenamefont {Yuan}, \citenamefont {Cao}, \citenamefont {Kamra}, \citenamefont {Duine},\ and\ \citenamefont {Yan}}]{Review_QuantumMagnonics}%
  \BibitemOpen
  \bibfield  {author} {\bibinfo {author} {\bibfnamefont {H.}~\bibnamefont {Yuan}}, \bibinfo {author} {\bibfnamefont {Y.}~\bibnamefont {Cao}}, \bibinfo {author} {\bibfnamefont {A.}~\bibnamefont {Kamra}}, \bibinfo {author} {\bibfnamefont {R.~A.}\ \bibnamefont {Duine}}, \ and\ \bibinfo {author} {\bibfnamefont {P.}~\bibnamefont {Yan}},\ }\bibfield  {title} {\enquote {\bibinfo {title} {Quantum magnonics: when magnon spintronics meets quantum information science},}\ }\href {https://doi.org/10.1016/j.physrep.2022.03.002} {\bibfield  {journal} {\bibinfo  {journal} {Phys. Rep.}\ }\textbf {\bibinfo {volume} {965}},\ \bibinfo {pages} {1} (\bibinfo {year} {2022})},\ \Eprint {http://arxiv.org/abs/arXiv:2111.14241} {arXiv:2111.14241} \BibitemShut {NoStop}%
\bibitem [{\citenamefont {Nakata}\ \emph {et~al.}(2014)\citenamefont {Nakata}, \citenamefont {van Hoogdalem}, \citenamefont {Simon},\ and\ \citenamefont {Loss}}]{KKPD}%
  \BibitemOpen
  \bibfield  {author} {\bibinfo {author} {\bibfnamefont {K.}~\bibnamefont {Nakata}}, \bibinfo {author} {\bibfnamefont {K.~A.}\ \bibnamefont {van Hoogdalem}}, \bibinfo {author} {\bibfnamefont {P.}~\bibnamefont {Simon}}, \ and\ \bibinfo {author} {\bibfnamefont {D.}~\bibnamefont {Loss}},\ }\bibfield  {title} {\enquote {\bibinfo {title} {Josephson and persistent spin currents in {B}ose-{E}instein condensates of magnons},}\ }\href {https://doi.org/10.1103/PhysRevB.90.144419} {\bibfield  {journal} {\bibinfo  {journal} {Phys. Rev. B}\ }\textbf {\bibinfo {volume} {90}},\ \bibinfo {pages} {144419} (\bibinfo {year} {2014})},\ \Eprint {http://arxiv.org/abs/arXiv:1406.7004} {arXiv:1406.7004} \BibitemShut {NoStop}%
\bibitem [{\citenamefont {Troncoso}\ and\ \citenamefont {N{\'u}{\~n}ez}(2014)}]{TroncosoJosephson}%
  \BibitemOpen
  \bibfield  {author} {\bibinfo {author} {\bibfnamefont {R.~E.}\ \bibnamefont {Troncoso}}\ and\ \bibinfo {author} {\bibfnamefont {{\'A}.~S.}\ \bibnamefont {N{\'u}{\~n}ez}},\ }\bibfield  {title} {\enquote {\bibinfo {title} {Josephson effects in a {B}ose--{E}instein condensate of magnons},}\ }\href {https://doi.org/10.1016/j.aop.2014.04.017} {\bibfield  {journal} {\bibinfo  {journal} {Ann. Phys. (N. Y.)}\ }\textbf {\bibinfo {volume} {346}},\ \bibinfo {pages} {182} (\bibinfo {year} {2014})},\ \Eprint {http://arxiv.org/abs/arXiv:1305.4285} {arXiv:1305.4285} \BibitemShut {NoStop}%
\bibitem [{\citenamefont {Liu}\ \emph {et~al.}(2016)\citenamefont {Liu}, \citenamefont {Yin}, \citenamefont {Zang}, \citenamefont {Lake},\ and\ \citenamefont {Barlas}}]{AFJosephsonTHz}%
  \BibitemOpen
  \bibfield  {author} {\bibinfo {author} {\bibfnamefont {Y.}~\bibnamefont {Liu}}, \bibinfo {author} {\bibfnamefont {G.}~\bibnamefont {Yin}}, \bibinfo {author} {\bibfnamefont {J.}~\bibnamefont {Zang}}, \bibinfo {author} {\bibfnamefont {R.~K.}\ \bibnamefont {Lake}}, \ and\ \bibinfo {author} {\bibfnamefont {Y.}~\bibnamefont {Barlas}},\ }\bibfield  {title} {\enquote {\bibinfo {title} {Spin-{J}osephson effects in exchange coupled antiferromagnetic insulators},}\ }\href {https://doi.org/10.1103/PhysRevB.94.094434} {\bibfield  {journal} {\bibinfo  {journal} {Phys. Rev. B}\ }\textbf {\bibinfo {volume} {94}},\ \bibinfo {pages} {094434} (\bibinfo {year} {2016})},\ \Eprint {http://arxiv.org/abs/arXiv:1605.09427} {arXiv:1605.09427} \BibitemShut {NoStop}%
\bibitem [{\citenamefont {Khymyn}\ \emph {et~al.}(2017)\citenamefont {Khymyn}, \citenamefont {Lisenkov}, \citenamefont {Tiberkevich}, \citenamefont {Ivanov},\ and\ \citenamefont {Slavin}}]{AFJosephsonTHz2}%
  \BibitemOpen
  \bibfield  {author} {\bibinfo {author} {\bibfnamefont {R.}~\bibnamefont {Khymyn}}, \bibinfo {author} {\bibfnamefont {I.}~\bibnamefont {Lisenkov}}, \bibinfo {author} {\bibfnamefont {V.}~\bibnamefont {Tiberkevich}}, \bibinfo {author} {\bibfnamefont {B.~A.}\ \bibnamefont {Ivanov}}, \ and\ \bibinfo {author} {\bibfnamefont {A.}~\bibnamefont {Slavin}},\ }\bibfield  {title} {\enquote {\bibinfo {title} {Antiferromagnetic {TH}z-frequency {J}osephson-like oscillator driven by spin current},}\ }\href {https://www.nature.com/articles/srep43705#citeas} {\bibfield  {journal} {\bibinfo  {journal} {Sci. Rep.}\ }\textbf {\bibinfo {volume} {7}},\ \bibinfo {pages} {43705} (\bibinfo {year} {2017})},\ \Eprint {http://arxiv.org/abs/arXiv:1609.09866} {arXiv:1609.09866} \BibitemShut {NoStop}%
\bibitem [{\citenamefont {Yu}\ \emph {et~al.}(2024)\citenamefont {Yu}, \citenamefont {Zou}, \citenamefont {Zeng}, \citenamefont {Rao},\ and\ \citenamefont {Xia}}]{ReviewNHmagnonics}%
  \BibitemOpen
  \bibfield  {author} {\bibinfo {author} {\bibfnamefont {T.}~\bibnamefont {Yu}}, \bibinfo {author} {\bibfnamefont {J.}~\bibnamefont {Zou}}, \bibinfo {author} {\bibfnamefont {B.}~\bibnamefont {Zeng}}, \bibinfo {author} {\bibfnamefont {J.}~\bibnamefont {Rao}}, \ and\ \bibinfo {author} {\bibfnamefont {K.}~\bibnamefont {Xia}},\ }\bibfield  {title} {\enquote {\bibinfo {title} {Non-{H}ermitian topological magnonics},}\ }\href {https://doi.org/10.1016/j.physrep.2024.01.006} {\bibfield  {journal} {\bibinfo  {journal} {Phys. Rep.}\ }\textbf {\bibinfo {volume} {1062}},\ \bibinfo {pages} {1} (\bibinfo {year} {2024})},\ \Eprint {http://arxiv.org/abs/arXiv:2306.04348} {arXiv:2306.04348} \BibitemShut {NoStop}%
\bibitem [{\citenamefont {Tserkovnyak}(2020)}]{YT_EP_LLG}%
  \BibitemOpen
  \bibfield  {author} {\bibinfo {author} {\bibfnamefont {Y.}~\bibnamefont {Tserkovnyak}},\ }\bibfield  {title} {\enquote {\bibinfo {title} {Exceptional points in dissipatively coupled spin dynamics},}\ }\href {https://doi.org/10.1103/PhysRevResearch.2.013031} {\bibfield  {journal} {\bibinfo  {journal} {Phys. Rev. Res.}\ }\textbf {\bibinfo {volume} {2}},\ \bibinfo {pages} {013031} (\bibinfo {year} {2020})},\ \Eprint {http://arxiv.org/abs/arXiv:1911.01619} {arXiv:1911.01619} \BibitemShut {NoStop}%
\bibitem [{\citenamefont {Zou}\ \emph {et~al.}(2024{\natexlab{a}})\citenamefont {Zou}, \citenamefont {Bosco}, \citenamefont {Thingstad}, \citenamefont {Klinovaja},\ and\ \citenamefont {Loss}}]{Ji_NHdiode}%
  \BibitemOpen
  \bibfield  {author} {\bibinfo {author} {\bibfnamefont {J.}~\bibnamefont {Zou}}, \bibinfo {author} {\bibfnamefont {S.}~\bibnamefont {Bosco}}, \bibinfo {author} {\bibfnamefont {E.}~\bibnamefont {Thingstad}}, \bibinfo {author} {\bibfnamefont {J.}~\bibnamefont {Klinovaja}}, \ and\ \bibinfo {author} {\bibfnamefont {D.}~\bibnamefont {Loss}},\ }\bibfield  {title} {\enquote {\bibinfo {title} {Dissipative spin-wave diode and nonreciprocal magnonic amplifier},}\ }\href {https://doi.org/10.1103/PhysRevLett.132.036701} {\bibfield  {journal} {\bibinfo  {journal} {Phys. Rev. Lett.}\ }\textbf {\bibinfo {volume} {132}},\ \bibinfo {pages} {036701} (\bibinfo {year} {2024}{\natexlab{a}})},\ \Eprint {http://arxiv.org/abs/arXiv:2306.15916} {arXiv:2306.15916} \BibitemShut {NoStop}%
\bibitem [{\citenamefont {Bulaevskii}\ \emph {et~al.}(1977)\citenamefont {Bulaevskii}, \citenamefont {Kuzii},\ and\ \citenamefont {Sobyanin}}]{piTheory1977}%
  \BibitemOpen
  \bibfield  {author} {\bibinfo {author} {\bibfnamefont {L.}~\bibnamefont {Bulaevskii}}, \bibinfo {author} {\bibfnamefont {V.}~\bibnamefont {Kuzii}}, \ and\ \bibinfo {author} {\bibfnamefont {A.}~\bibnamefont {Sobyanin}},\ }\bibfield  {title} {\enquote {\bibinfo {title} {Superconducting system with weak coupling to the current in the ground state},}\ }\href {http://www.jetpletters.ru/ps/1410/article_21163.shtml} {\bibfield  {journal} {\bibinfo  {journal} {J. Exp. Theor. Phys. Lett.}\ }\textbf {\bibinfo {volume} {25}},\ \bibinfo {pages} {290} (\bibinfo {year} {1977})}\BibitemShut {NoStop}%
\bibitem [{\citenamefont {Geshkenbein}\ \emph {et~al.}(1987)\citenamefont {Geshkenbein}, \citenamefont {Larkin},\ and\ \citenamefont {Barone}}]{piTheory1987}%
  \BibitemOpen
  \bibfield  {author} {\bibinfo {author} {\bibfnamefont {V.~B.}\ \bibnamefont {Geshkenbein}}, \bibinfo {author} {\bibfnamefont {A.~I.}\ \bibnamefont {Larkin}}, \ and\ \bibinfo {author} {\bibfnamefont {A.}~\bibnamefont {Barone}},\ }\bibfield  {title} {\enquote {\bibinfo {title} {Vortices with half magnetic flux quanta in ``heavy-fermion'' superconductors},}\ }\href {https://doi.org/10.1103/PhysRevB.36.235} {\bibfield  {journal} {\bibinfo  {journal} {Phys. Rev. B}\ }\textbf {\bibinfo {volume} {36}},\ \bibinfo {pages} {235} (\bibinfo {year} {1987})}\BibitemShut {NoStop}%
\bibitem [{\citenamefont {Sigrist}\ and\ \citenamefont {Rice}(1992)}]{piTheory1992}%
  \BibitemOpen
  \bibfield  {author} {\bibinfo {author} {\bibfnamefont {M.}~\bibnamefont {Sigrist}}\ and\ \bibinfo {author} {\bibfnamefont {T.~M.}\ \bibnamefont {Rice}},\ }\bibfield  {title} {\enquote {\bibinfo {title} {Paramagnetic effect in high ${T}_{\text{c}}$ superconductors-a hint for $d$-wave superconductivity},}\ }\href {https://doi.org/10.1143/JPSJ.61.4283} {\bibfield  {journal} {\bibinfo  {journal} {J. Phys. Soc. Jpn.}\ }\textbf {\bibinfo {volume} {61}},\ \bibinfo {pages} {4283} (\bibinfo {year} {1992})}\BibitemShut {NoStop}%
\bibitem [{\citenamefont {Wollman}\ \emph {et~al.}(1993)\citenamefont {Wollman}, \citenamefont {Van~Harlingen}, \citenamefont {Lee}, \citenamefont {Ginsberg},\ and\ \citenamefont {Leggett}}]{piExp1993}%
  \BibitemOpen
  \bibfield  {author} {\bibinfo {author} {\bibfnamefont {D.~A.}\ \bibnamefont {Wollman}}, \bibinfo {author} {\bibfnamefont {D.~J.}\ \bibnamefont {Van~Harlingen}}, \bibinfo {author} {\bibfnamefont {W.~C.}\ \bibnamefont {Lee}}, \bibinfo {author} {\bibfnamefont {D.~M.}\ \bibnamefont {Ginsberg}}, \ and\ \bibinfo {author} {\bibfnamefont {A.~J.}\ \bibnamefont {Leggett}},\ }\bibfield  {title} {\enquote {\bibinfo {title} {Experimental determination of the superconducting pairing state in {YBCO} from the phase coherence of {YBCO}-{P}b dc {SQUID}s},}\ }\href {https://doi.org/10.1103/PhysRevLett.71.2134} {\bibfield  {journal} {\bibinfo  {journal} {Phys. Rev. Lett.}\ }\textbf {\bibinfo {volume} {71}},\ \bibinfo {pages} {2134} (\bibinfo {year} {1993})}\BibitemShut {NoStop}%
\bibitem [{\citenamefont {Van~Harlingen}(1995)}]{piExp1995}%
  \BibitemOpen
  \bibfield  {author} {\bibinfo {author} {\bibfnamefont {D.~J.}\ \bibnamefont {Van~Harlingen}},\ }\bibfield  {title} {\enquote {\bibinfo {title} {Phase-sensitive tests of the symmetry of the pairing state in the high-temperature superconductors---{E}vidence for ${d}_{{x}^{2}\ensuremath{-}{y}^{2}}$ symmetry},}\ }\href {https://doi.org/10.1103/RevModPhys.67.515} {\bibfield  {journal} {\bibinfo  {journal} {Rev. Mod. Phys.}\ }\textbf {\bibinfo {volume} {67}},\ \bibinfo {pages} {515} (\bibinfo {year} {1995})}\BibitemShut {NoStop}%
\bibitem [{\citenamefont {Ryazanov}\ \emph {et~al.}(2001)\citenamefont {Ryazanov}, \citenamefont {Oboznov}, \citenamefont {Rusanov}, \citenamefont {Veretennikov}, \citenamefont {Golubov},\ and\ \citenamefont {Aarts}}]{piExp2001}%
  \BibitemOpen
  \bibfield  {author} {\bibinfo {author} {\bibfnamefont {V.~V.}\ \bibnamefont {Ryazanov}}, \bibinfo {author} {\bibfnamefont {V.~A.}\ \bibnamefont {Oboznov}}, \bibinfo {author} {\bibfnamefont {A.~Yu.}\ \bibnamefont {Rusanov}}, \bibinfo {author} {\bibfnamefont {A.~V.}\ \bibnamefont {Veretennikov}}, \bibinfo {author} {\bibfnamefont {A.~A.}\ \bibnamefont {Golubov}}, \ and\ \bibinfo {author} {\bibfnamefont {J.}~\bibnamefont {Aarts}},\ }\bibfield  {title} {\enquote {\bibinfo {title} {Coupling of two superconductors through a ferromagnet: Evidence for a $\ensuremath{\pi}$ junction},}\ }\href {https://doi.org/10.1103/PhysRevLett.86.2427} {\bibfield  {journal} {\bibinfo  {journal} {Phys. Rev. Lett.}\ }\textbf {\bibinfo {volume} {86}},\ \bibinfo {pages} {2427} (\bibinfo {year} {2001})},\ \Eprint {http://arxiv.org/abs/arXiv:cond-mat/0008364} {arXiv:cond-mat/0008364} \BibitemShut {NoStop}%
\bibitem [{\citenamefont {Il'ichev}\ \emph {et~al.}(2001)\citenamefont {Il'ichev}, \citenamefont {Grajcar}, \citenamefont {Hlubina}, \citenamefont {IJsselsteijn}, \citenamefont {Hoenig}, \citenamefont {Meyer}, \citenamefont {Golubov}, \citenamefont {Amin}, \citenamefont {Zagoskin}, \citenamefont {Omelyanchouk},\ and\ \citenamefont {Kupriyanov}}]{phiExp2001}%
  \BibitemOpen
  \bibfield  {author} {\bibinfo {author} {\bibfnamefont {E.}~\bibnamefont {Il'ichev}}, \bibinfo {author} {\bibfnamefont {M.}~\bibnamefont {Grajcar}}, \bibinfo {author} {\bibfnamefont {R.}~\bibnamefont {Hlubina}}, \bibinfo {author} {\bibfnamefont {R.~P.~J.}\ \bibnamefont {IJsselsteijn}}, \bibinfo {author} {\bibfnamefont {H.~E.}\ \bibnamefont {Hoenig}}, \bibinfo {author} {\bibfnamefont {H.-G.}\ \bibnamefont {Meyer}}, \bibinfo {author} {\bibfnamefont {A.}~\bibnamefont {Golubov}}, \bibinfo {author} {\bibfnamefont {M.~H.~S.}\ \bibnamefont {Amin}}, \bibinfo {author} {\bibfnamefont {A.~M.}\ \bibnamefont {Zagoskin}}, \bibinfo {author} {\bibfnamefont {A.~N.}\ \bibnamefont {Omelyanchouk}}, \ and\ \bibinfo {author} {\bibfnamefont {M.~Yu.}\ \bibnamefont {Kupriyanov}},\ }\bibfield  {title} {\enquote {\bibinfo {title} {Degenerate ground state in a mesoscopic ${{\mathrm{{YB}a}}_{2}{\mathrm{{C}u}}_{3} \mathrm{O}}_{7\ensuremath{-}\mathit{x}}$ grain boundary {J}osephson junction},}\ }\href {https://doi.org/10.1103/PhysRevLett.86.5369} {\bibfield  {journal} {\bibinfo  {journal} {Phys. Rev. Lett.}\ }\textbf {\bibinfo {volume} {86}},\ \bibinfo {pages} {5369} (\bibinfo {year} {2001})},\ \Eprint {http://arxiv.org/abs/arXiv:cond-mat/0102404} {arXiv:cond-mat/0102404} \BibitemShut {NoStop}%
\bibitem [{\citenamefont {Testa}\ \emph {et~al.}(2004)\citenamefont {Testa}, \citenamefont {Monaco}, \citenamefont {Esposito}, \citenamefont {Sarnelli}, \citenamefont {Kang}, \citenamefont {Mennema}, \citenamefont {Tarte},\ and\ \citenamefont {Blamire}}]{phiExp2004}%
  \BibitemOpen
  \bibfield  {author} {\bibinfo {author} {\bibfnamefont {G.}~\bibnamefont {Testa}}, \bibinfo {author} {\bibfnamefont {A.}~\bibnamefont {Monaco}}, \bibinfo {author} {\bibfnamefont {E.}~\bibnamefont {Esposito}}, \bibinfo {author} {\bibfnamefont {E.}~\bibnamefont {Sarnelli}}, \bibinfo {author} {\bibfnamefont {D.-J.}\ \bibnamefont {Kang}}, \bibinfo {author} {\bibfnamefont {S.}~\bibnamefont {Mennema}}, \bibinfo {author} {\bibfnamefont {E.}~\bibnamefont {Tarte}}, \ and\ \bibinfo {author} {\bibfnamefont {M.}~\bibnamefont {Blamire}},\ }\bibfield  {title} {\enquote {\bibinfo {title} {Midgap state-based $\pi$-junctions for digital applications},}\ }\href {https://doi.org/10.1063/1.1781744} {\bibfield  {journal} {\bibinfo  {journal} {Appl. Phys. Lett.}\ }\textbf {\bibinfo {volume} {85}},\ \bibinfo {pages} {1202} (\bibinfo {year} {2004})}\BibitemShut {NoStop}%
\bibitem [{\citenamefont {Sickinger}\ \emph {et~al.}(2012)\citenamefont {Sickinger}, \citenamefont {Lipman}, \citenamefont {Weides}, \citenamefont {Mints}, \citenamefont {Kohlstedt}, \citenamefont {Koelle}, \citenamefont {Kleiner},\ and\ \citenamefont {Goldobin}}]{phiExp2012}%
  \BibitemOpen
  \bibfield  {author} {\bibinfo {author} {\bibfnamefont {H.}~\bibnamefont {Sickinger}}, \bibinfo {author} {\bibfnamefont {A.}~\bibnamefont {Lipman}}, \bibinfo {author} {\bibfnamefont {M.}~\bibnamefont {Weides}}, \bibinfo {author} {\bibfnamefont {R.~G.}\ \bibnamefont {Mints}}, \bibinfo {author} {\bibfnamefont {H.}~\bibnamefont {Kohlstedt}}, \bibinfo {author} {\bibfnamefont {D.}~\bibnamefont {Koelle}}, \bibinfo {author} {\bibfnamefont {R.}~\bibnamefont {Kleiner}}, \ and\ \bibinfo {author} {\bibfnamefont {E.}~\bibnamefont {Goldobin}},\ }\bibfield  {title} {\enquote {\bibinfo {title} {Experimental evidence of a $\ensuremath{\varphi}$ {J}osephson junction},}\ }\href {https://doi.org/10.1103/PhysRevLett.109.107002} {\bibfield  {journal} {\bibinfo  {journal} {Phys. Rev. Lett.}\ }\textbf {\bibinfo {volume} {109}},\ \bibinfo {pages} {107002} (\bibinfo {year} {2012})},\ \Eprint {http://arxiv.org/abs/arXiv:1207.3013} {arXiv:1207.3013} \BibitemShut {NoStop}%
\bibitem [{\citenamefont {Goldobin}\ \emph {et~al.}(2013)\citenamefont {Goldobin}, \citenamefont {Sickinger}, \citenamefont {Weides}, \citenamefont {Ruppelt}, \citenamefont {Kohlstedt}, \citenamefont {Kleiner},\ and\ \citenamefont {Koelle}}]{phiExp2013}%
  \BibitemOpen
  \bibfield  {author} {\bibinfo {author} {\bibfnamefont {E.}~\bibnamefont {Goldobin}}, \bibinfo {author} {\bibfnamefont {H.}~\bibnamefont {Sickinger}}, \bibinfo {author} {\bibfnamefont {M.}~\bibnamefont {Weides}}, \bibinfo {author} {\bibfnamefont {N.}~\bibnamefont {Ruppelt}}, \bibinfo {author} {\bibfnamefont {H.}~\bibnamefont {Kohlstedt}}, \bibinfo {author} {\bibfnamefont {R.}~\bibnamefont {Kleiner}}, \ and\ \bibinfo {author} {\bibfnamefont {D.}~\bibnamefont {Koelle}},\ }\bibfield  {title} {\enquote {\bibinfo {title} {Memory cell based on a $\varphi$ {J}osephson junction},}\ }\href {https://doi.org/10.1063/1.4811752} {\bibfield  {journal} {\bibinfo  {journal} {Appl. Phys. Lett.}\ }\textbf {\bibinfo {volume} {102}},\ \bibinfo {pages} {242602} (\bibinfo {year} {2013})},\ \Eprint {http://arxiv.org/abs/arXiv:1306.1683} {arXiv:1306.1683} \BibitemShut {NoStop}%
\bibitem [{\citenamefont {Szombati}\ \emph {et~al.}(2016)\citenamefont {Szombati}, \citenamefont {Nadj-Perge}, \citenamefont {Car}, \citenamefont {Plissard}, \citenamefont {Bakkers},\ and\ \citenamefont {Kouwenhoven}}]{phiExp2016}%
  \BibitemOpen
  \bibfield  {author} {\bibinfo {author} {\bibfnamefont {D.}~\bibnamefont {Szombati}}, \bibinfo {author} {\bibfnamefont {S.}~\bibnamefont {Nadj-Perge}}, \bibinfo {author} {\bibfnamefont {D.}~\bibnamefont {Car}}, \bibinfo {author} {\bibfnamefont {S.}~\bibnamefont {Plissard}}, \bibinfo {author} {\bibfnamefont {E.}~\bibnamefont {Bakkers}}, \ and\ \bibinfo {author} {\bibfnamefont {L.}~\bibnamefont {Kouwenhoven}},\ }\bibfield  {title} {\enquote {\bibinfo {title} {Josephson $\varphi_0$-junction in nanowire quantum dots},}\ }\href {https://doi.org/10.1038/nphys3742} {\bibfield  {journal} {\bibinfo  {journal} {Nat. Phys.}\ }\textbf {\bibinfo {volume} {12}},\ \bibinfo {pages} {568} (\bibinfo {year} {2016})},\ \Eprint {http://arxiv.org/abs/arXiv:1512.01234} {arXiv:1512.01234} \BibitemShut {NoStop}%
\bibitem [{\citenamefont {Tanaka}\ and\ \citenamefont {Kashiwaya}(1996)}]{phiTheory1996}%
  \BibitemOpen
  \bibfield  {author} {\bibinfo {author} {\bibfnamefont {Y.}~\bibnamefont {Tanaka}}\ and\ \bibinfo {author} {\bibfnamefont {S.}~\bibnamefont {Kashiwaya}},\ }\bibfield  {title} {\enquote {\bibinfo {title} {Theory of the {J}osephson effect in $d$-wave superconductors},}\ }\href {https://doi.org/10.1103/PhysRevB.53.R11957} {\bibfield  {journal} {\bibinfo  {journal} {Phys. Rev. B}\ }\textbf {\bibinfo {volume} {53}},\ \bibinfo {pages} {R11957} (\bibinfo {year} {1996})}\BibitemShut {NoStop}%
\bibitem [{\citenamefont {Barash}\ \emph {et~al.}(1996)\citenamefont {Barash}, \citenamefont {Burkhardt},\ and\ \citenamefont {Rainer}}]{phiTheory1996_2}%
  \BibitemOpen
  \bibfield  {author} {\bibinfo {author} {\bibfnamefont {Y.~S.}\ \bibnamefont {Barash}}, \bibinfo {author} {\bibfnamefont {H.}~\bibnamefont {Burkhardt}}, \ and\ \bibinfo {author} {\bibfnamefont {D.}~\bibnamefont {Rainer}},\ }\bibfield  {title} {\enquote {\bibinfo {title} {Low-temperature anomaly in the {J}osephson critical current of junctions in $\mathit{d}$-wave superconductors},}\ }\href {https://doi.org/10.1103/PhysRevLett.77.4070} {\bibfield  {journal} {\bibinfo  {journal} {Phys. Rev. Lett.}\ }\textbf {\bibinfo {volume} {77}},\ \bibinfo {pages} {4070} (\bibinfo {year} {1996})}\BibitemShut {NoStop}%
\bibitem [{\citenamefont {Tanaka}\ and\ \citenamefont {Kashiwaya}(1997)}]{phiTheory1997}%
  \BibitemOpen
  \bibfield  {author} {\bibinfo {author} {\bibfnamefont {Y.}~\bibnamefont {Tanaka}}\ and\ \bibinfo {author} {\bibfnamefont {S.}~\bibnamefont {Kashiwaya}},\ }\bibfield  {title} {\enquote {\bibinfo {title} {Theory of {J}osephson effects in anisotropic superconductors},}\ }\href {https://doi.org/10.1103/PhysRevB.56.892} {\bibfield  {journal} {\bibinfo  {journal} {Phys. Rev. B}\ }\textbf {\bibinfo {volume} {56}},\ \bibinfo {pages} {892} (\bibinfo {year} {1997})}\BibitemShut {NoStop}%
\bibitem [{\citenamefont {Mints}(1998)}]{phiTheory1998}%
  \BibitemOpen
  \bibfield  {author} {\bibinfo {author} {\bibfnamefont {R.~G.}\ \bibnamefont {Mints}},\ }\bibfield  {title} {\enquote {\bibinfo {title} {Self-generated flux in {J}osephson junctions with alternating critical current density},}\ }\href {https://doi.org/10.1103/PhysRevB.57.R3221} {\bibfield  {journal} {\bibinfo  {journal} {Phys. Rev. B}\ }\textbf {\bibinfo {volume} {57}},\ \bibinfo {pages} {R3221} (\bibinfo {year} {1998})}\BibitemShut {NoStop}%
\bibitem [{\citenamefont {Mints}\ and\ \citenamefont {Papiashvili}(2001)}]{phiTheory2001}%
  \BibitemOpen
  \bibfield  {author} {\bibinfo {author} {\bibfnamefont {R.~G.}\ \bibnamefont {Mints}}\ and\ \bibinfo {author} {\bibfnamefont {I.}~\bibnamefont {Papiashvili}},\ }\bibfield  {title} {\enquote {\bibinfo {title} {Josephson vortices with fractional flux quanta at ${{\mathrm{{YB}a}}}_{2}{{{\mathrm{{C}}}}\mathrm{u}}_{3}{{\mathrm{{O}}}}_{7\ensuremath{-}x}$ grain boundaries},}\ }\href {https://doi.org/10.1103/PhysRevB.64.134501} {\bibfield  {journal} {\bibinfo  {journal} {Phys. Rev. B}\ }\textbf {\bibinfo {volume} {64}},\ \bibinfo {pages} {134501} (\bibinfo {year} {2001})},\ \Eprint {http://arxiv.org/abs/arXiv:cond-mat/0101085} {arXiv:cond-mat/0101085} \BibitemShut {NoStop}%
\bibitem [{\citenamefont {Buzdin}\ and\ \citenamefont {Koshelev}(2003)}]{phiTheory2003}%
  \BibitemOpen
  \bibfield  {author} {\bibinfo {author} {\bibfnamefont {A.}~\bibnamefont {Buzdin}}\ and\ \bibinfo {author} {\bibfnamefont {A.~E.}\ \bibnamefont {Koshelev}},\ }\bibfield  {title} {\enquote {\bibinfo {title} {Periodic alternating $0$- and $\ensuremath{\pi}$-junction structures as realization of $\ensuremath{\varphi}$-{J}osephson junctions},}\ }\href {https://doi.org/10.1103/PhysRevB.67.220504} {\bibfield  {journal} {\bibinfo  {journal} {Phys. Rev. B}\ }\textbf {\bibinfo {volume} {67}},\ \bibinfo {pages} {220504} (\bibinfo {year} {2003})},\ \Eprint {http://arxiv.org/abs/arXiv:cond-mat/0305142} {arXiv:cond-mat/0305142} \BibitemShut {NoStop}%
\bibitem [{\citenamefont {Gumann}\ \emph {et~al.}(2007)\citenamefont {Gumann}, \citenamefont {Iniotakis},\ and\ \citenamefont {Schopohl}}]{phiTheory2007}%
  \BibitemOpen
  \bibfield  {author} {\bibinfo {author} {\bibfnamefont {A.}~\bibnamefont {Gumann}}, \bibinfo {author} {\bibfnamefont {C.}~\bibnamefont {Iniotakis}}, \ and\ \bibinfo {author} {\bibfnamefont {N.}~\bibnamefont {Schopohl}},\ }\bibfield  {title} {\enquote {\bibinfo {title} {Geometric $\pi$ {J}osephson junction in $d$-wave superconducting thin films},}\ }\href {https://doi.org/10.1063/1.2801387} {\bibfield  {journal} {\bibinfo  {journal} {Appl. Phys. Lett.}\ }\textbf {\bibinfo {volume} {91}},\ \bibinfo {pages} {192502} (\bibinfo {year} {2007})},\ \Eprint {http://arxiv.org/abs/arXiv:0708.3898} {arXiv:0708.3898} \BibitemShut {NoStop}%
\bibitem [{\citenamefont {Buzdin}(2008)}]{PhiTheory2008}%
  \BibitemOpen
  \bibfield  {author} {\bibinfo {author} {\bibfnamefont {A.}~\bibnamefont {Buzdin}},\ }\bibfield  {title} {\enquote {\bibinfo {title} {Direct coupling between magnetism and superconducting current in the {J}osephson ${\ensuremath{\varphi}}_{0}$ junction},}\ }\href {https://doi.org/10.1103/PhysRevLett.101.107005} {\bibfield  {journal} {\bibinfo  {journal} {Phys. Rev. Lett.}\ }\textbf {\bibinfo {volume} {101}},\ \bibinfo {pages} {107005} (\bibinfo {year} {2008})},\ \Eprint {http://arxiv.org/abs/arXiv:0808.0299} {arXiv:0808.0299} \BibitemShut {NoStop}%
\bibitem [{\citenamefont {Goldobin}\ \emph {et~al.}(2011)\citenamefont {Goldobin}, \citenamefont {Koelle}, \citenamefont {Kleiner},\ and\ \citenamefont {Mints}}]{phiTheory2011}%
  \BibitemOpen
  \bibfield  {author} {\bibinfo {author} {\bibfnamefont {E.}~\bibnamefont {Goldobin}}, \bibinfo {author} {\bibfnamefont {D.}~\bibnamefont {Koelle}}, \bibinfo {author} {\bibfnamefont {R.}~\bibnamefont {Kleiner}}, \ and\ \bibinfo {author} {\bibfnamefont {R.~G.}\ \bibnamefont {Mints}},\ }\bibfield  {title} {\enquote {\bibinfo {title} {Josephson junction with a magnetic-field tunable ground state},}\ }\href {https://doi.org/10.1103/PhysRevLett.107.227001} {\bibfield  {journal} {\bibinfo  {journal} {Phys. Rev. Lett.}\ }\textbf {\bibinfo {volume} {107}},\ \bibinfo {pages} {227001} (\bibinfo {year} {2011})},\ \Eprint {http://arxiv.org/abs/arXiv:1110.2326} {arXiv:1110.2326} \BibitemShut {NoStop}%
\bibitem [{Note1()}]{Note1}%
  \BibitemOpen
  \bibinfo {note} {See Ref.~\cite {AJE_exp_2023} for observation of anomalous Josephson effect in metallic and nonmagnetic superconducting-normal-superconducting junctions configured as nonequilibrium Andreev interferometers~\cite {AJE_Theory_2019,AJE_Theory_2019PRB,AJE_Theory_2021}.}\BibitemShut {Stop}%
\bibitem [{\citenamefont {Nakata}\ \emph {et~al.}(2015{\natexlab{a}})\citenamefont {Nakata}, \citenamefont {Simon},\ and\ \citenamefont {Loss}}]{KPD}%
  \BibitemOpen
  \bibfield  {author} {\bibinfo {author} {\bibfnamefont {K.}~\bibnamefont {Nakata}}, \bibinfo {author} {\bibfnamefont {P.}~\bibnamefont {Simon}}, \ and\ \bibinfo {author} {\bibfnamefont {D.}~\bibnamefont {Loss}},\ }\bibfield  {title} {\enquote {\bibinfo {title} {Magnon transport through microwave pumping},}\ }\href {https://doi.org/10.1103/PhysRevB.92.014422} {\bibfield  {journal} {\bibinfo  {journal} {Phys. Rev. B}\ }\textbf {\bibinfo {volume} {92}},\ \bibinfo {pages} {014422} (\bibinfo {year} {2015}{\natexlab{a}})},\ \Eprint {http://arxiv.org/abs/arXiv:1502.03865} {arXiv:1502.03865} \BibitemShut {NoStop}%
\bibitem [{\citenamefont {Holstein}\ and\ \citenamefont {Primakoff}(1940)}]{HP}%
  \BibitemOpen
  \bibfield  {author} {\bibinfo {author} {\bibfnamefont {T.}~\bibnamefont {Holstein}}\ and\ \bibinfo {author} {\bibfnamefont {H.}~\bibnamefont {Primakoff}},\ }\bibfield  {title} {\enquote {\bibinfo {title} {Field dependence of the intrinsic domain magnetization of a ferromagnet},}\ }\href {https://doi.org/10.1103/PhysRev.58.1098} {\bibfield  {journal} {\bibinfo  {journal} {Phys. Rev.}\ }\textbf {\bibinfo {volume} {58}},\ \bibinfo {pages} {1098} (\bibinfo {year} {1940})}\BibitemShut {NoStop}%
\bibitem [{\citenamefont {Meier}\ and\ \citenamefont {Loss}(2003)}]{magnon2}%
  \BibitemOpen
  \bibfield  {author} {\bibinfo {author} {\bibfnamefont {F.}~\bibnamefont {Meier}}\ and\ \bibinfo {author} {\bibfnamefont {D.}~\bibnamefont {Loss}},\ }\bibfield  {title} {\enquote {\bibinfo {title} {Magnetization transport and quantized spin conductance},}\ }\href {https://doi.org/10.1103/PhysRevLett.90.167204} {\bibfield  {journal} {\bibinfo  {journal} {Phys. Rev. Lett.}\ }\textbf {\bibinfo {volume} {90}},\ \bibinfo {pages} {167204} (\bibinfo {year} {2003})},\ \Eprint {http://arxiv.org/abs/arXiv:cond-mat/0209521} {arXiv:cond-mat/0209521} \BibitemShut {NoStop}%
\bibitem [{\citenamefont {Serha}\ \emph {et~al.}(2023)\citenamefont {Serha}, \citenamefont {Vasyuchka}, \citenamefont {Serga},\ and\ \citenamefont {Hillebrands}}]{ACphaseExp2023}%
  \BibitemOpen
  \bibfield  {author} {\bibinfo {author} {\bibfnamefont {R.~O.}\ \bibnamefont {Serha}}, \bibinfo {author} {\bibfnamefont {V.~I.}\ \bibnamefont {Vasyuchka}}, \bibinfo {author} {\bibfnamefont {A.~A.}\ \bibnamefont {Serga}}, \ and\ \bibinfo {author} {\bibfnamefont {B.}~\bibnamefont {Hillebrands}},\ }\bibfield  {title} {\enquote {\bibinfo {title} {Towards an experimental proof of the magnonic {A}haronov-{C}asher effect},}\ }\href {https://doi.org/10.1103/PhysRevB.108.L220404} {\bibfield  {journal} {\bibinfo  {journal} {Phys. Rev. B}\ }\textbf {\bibinfo {volume} {108}},\ \bibinfo {pages} {L220404} (\bibinfo {year} {2023})},\ \Eprint {http://arxiv.org/abs/arXiv:2312.05113} {arXiv:2312.05113} \BibitemShut {NoStop}%
\bibitem [{\citenamefont {Katsura}\ \emph {et~al.}(2005)\citenamefont {Katsura}, \citenamefont {Nagaosa},\ and\ \citenamefont {Balatsky}}]{katsura2}%
  \BibitemOpen
  \bibfield  {author} {\bibinfo {author} {\bibfnamefont {H.}~\bibnamefont {Katsura}}, \bibinfo {author} {\bibfnamefont {N.}~\bibnamefont {Nagaosa}}, \ and\ \bibinfo {author} {\bibfnamefont {A.~V.}\ \bibnamefont {Balatsky}},\ }\bibfield  {title} {\enquote {\bibinfo {title} {Spin current and magnetoelectric effect in noncollinear magnets},}\ }\href {https://doi.org/10.1103/PhysRevLett.95.057205} {\bibfield  {journal} {\bibinfo  {journal} {Phys. Rev. Lett.}\ }\textbf {\bibinfo {volume} {95}},\ \bibinfo {pages} {057205} (\bibinfo {year} {2005})},\ \Eprint {http://arxiv.org/abs/arXiv:cond-mat/0412319} {arXiv:cond-mat/0412319} \BibitemShut {NoStop}%
\bibitem [{\citenamefont {Liang}\ \emph {et~al.}(2023)\citenamefont {Liang}, \citenamefont {Chen}, \citenamefont {Cui}, \citenamefont {Zhou}, \citenamefont {Pan}, \citenamefont {Yang},\ and\ \citenamefont {Song}}]{ImJ2023MHz}%
  \BibitemOpen
  \bibfield  {author} {\bibinfo {author} {\bibfnamefont {S.}~\bibnamefont {Liang}}, \bibinfo {author} {\bibfnamefont {R.}~\bibnamefont {Chen}}, \bibinfo {author} {\bibfnamefont {Q.}~\bibnamefont {Cui}}, \bibinfo {author} {\bibfnamefont {Y.}~\bibnamefont {Zhou}}, \bibinfo {author} {\bibfnamefont {F.}~\bibnamefont {Pan}}, \bibinfo {author} {\bibfnamefont {H.}~\bibnamefont {Yang}}, \ and\ \bibinfo {author} {\bibfnamefont {C.}~\bibnamefont {Song}},\ }\bibfield  {title} {\enquote {\bibinfo {title} {{R}uderman--{K}ittel--{K}asuya--{Y}osida-type interlayer {D}zyaloshinskii--{M}oriya interaction in synthetic magnets},}\ }\href {https://doi.org/10.1021/acs.nanolett.3c02607} {\bibfield  {journal} {\bibinfo  {journal} {Nano Lett.}\ }\textbf {\bibinfo {volume} {23}},\ \bibinfo {pages} {8690} (\bibinfo {year} {2023})}\BibitemShut {NoStop}%
\bibitem [{\citenamefont {Kammerbauer}\ \emph {et~al.}(2023)\citenamefont {Kammerbauer}, \citenamefont {Choi}, \citenamefont {Freimuth}, \citenamefont {Lee}, \citenamefont {Fr$\ddot{\text{o}}$mter}, \citenamefont {Han}, \citenamefont {Lavrijsen}, \citenamefont {Swagten}, \citenamefont {Mokrousov},\ and\ \citenamefont {Kl$\ddot{\text{a}}$ui}}]{ImJ2023MHz2}%
  \BibitemOpen
  \bibfield  {author} {\bibinfo {author} {\bibfnamefont {F.}~\bibnamefont {Kammerbauer}}, \bibinfo {author} {\bibfnamefont {W.-Y.}\ \bibnamefont {Choi}}, \bibinfo {author} {\bibfnamefont {F.}~\bibnamefont {Freimuth}}, \bibinfo {author} {\bibfnamefont {K.}~\bibnamefont {Lee}}, \bibinfo {author} {\bibfnamefont {R.}~\bibnamefont {Fr$\ddot{\text{o}}$mter}}, \bibinfo {author} {\bibfnamefont {D.-S.}\ \bibnamefont {Han}}, \bibinfo {author} {\bibfnamefont {R.}~\bibnamefont {Lavrijsen}}, \bibinfo {author} {\bibfnamefont {H.~J.}\ \bibnamefont {Swagten}}, \bibinfo {author} {\bibfnamefont {Y.}~\bibnamefont {Mokrousov}}, \ and\ \bibinfo {author} {\bibfnamefont {M.}~\bibnamefont {Kl$\ddot{\text{a}}$ui}},\ }\bibfield  {title} {\enquote {\bibinfo {title} {Controlling the interlayer {D}zyaloshinskii--{M}oriya interaction by electrical currents},}\ }\href {https://doi.org/10.1021/acs.nanolett.3c01709} {\bibfield  {journal} {\bibinfo  {journal} {Nano Lett.}\ }\textbf {\bibinfo {volume} {23}},\ \bibinfo {pages} {7070} (\bibinfo {year} {2023})}\BibitemShut {NoStop}%
\bibitem [{\citenamefont {Heinrich}\ \emph {et~al.}(2003)\citenamefont {Heinrich}, \citenamefont {Tserkovnyak}, \citenamefont {Woltersdorf}, \citenamefont {Brataas}, \citenamefont {Urban},\ and\ \citenamefont {Bauer}}]{Damping_Heinrich2003}%
  \BibitemOpen
  \bibfield  {author} {\bibinfo {author} {\bibfnamefont {B.}~\bibnamefont {Heinrich}}, \bibinfo {author} {\bibfnamefont {Y.}~\bibnamefont {Tserkovnyak}}, \bibinfo {author} {\bibfnamefont {G.}~\bibnamefont {Woltersdorf}}, \bibinfo {author} {\bibfnamefont {A.}~\bibnamefont {Brataas}}, \bibinfo {author} {\bibfnamefont {R.}~\bibnamefont {Urban}}, \ and\ \bibinfo {author} {\bibfnamefont {G.~E.~W.}\ \bibnamefont {Bauer}},\ }\bibfield  {title} {\enquote {\bibinfo {title} {Dynamic exchange coupling in magnetic bilayers},}\ }\href {https://doi.org/10.1103/PhysRevLett.90.187601} {\bibfield  {journal} {\bibinfo  {journal} {Phys. Rev. Lett.}\ }\textbf {\bibinfo {volume} {90}},\ \bibinfo {pages} {187601} (\bibinfo {year} {2003})},\ \Eprint {http://arxiv.org/abs/arXiv:cond-mat/0210588} {arXiv:cond-mat/0210588} \BibitemShut {NoStop}%
\bibitem [{\citenamefont {Zou}\ \emph {et~al.}(2022)\citenamefont {Zou}, \citenamefont {Zhang},\ and\ \citenamefont {Tserkovnyak}}]{Ji_Dissipative_Bell}%
  \BibitemOpen
  \bibfield  {author} {\bibinfo {author} {\bibfnamefont {J.}~\bibnamefont {Zou}}, \bibinfo {author} {\bibfnamefont {S.}~\bibnamefont {Zhang}}, \ and\ \bibinfo {author} {\bibfnamefont {Y.}~\bibnamefont {Tserkovnyak}},\ }\bibfield  {title} {\enquote {\bibinfo {title} {Bell-state generation for spin qubits via dissipative coupling},}\ }\href {https://doi.org/10.1103/PhysRevB.106.L180406} {\bibfield  {journal} {\bibinfo  {journal} {Phys. Rev. B}\ }\textbf {\bibinfo {volume} {106}},\ \bibinfo {pages} {L180406} (\bibinfo {year} {2022})},\ \Eprint {http://arxiv.org/abs/arXiv:2108.07365} {arXiv:2108.07365} \BibitemShut {NoStop}%
\bibitem [{\citenamefont {Zou}\ \emph {et~al.}(2024{\natexlab{b}})\citenamefont {Zou}, \citenamefont {Bosco},\ and\ \citenamefont {Loss}}]{zou2023spatially}%
  \BibitemOpen
  \bibfield  {author} {\bibinfo {author} {\bibfnamefont {J.}~\bibnamefont {Zou}}, \bibinfo {author} {\bibfnamefont {S.}~\bibnamefont {Bosco}}, \ and\ \bibinfo {author} {\bibfnamefont {D.}~\bibnamefont {Loss}},\ }\bibfield  {title} {\enquote {\bibinfo {title} {Spatially correlated classical and quantum noise in driven qubits},}\ }\href {https://doi.org/10.1038/s41534-024-00842-9} {\bibfield  {journal} {\bibinfo  {journal} {npj Quantum Inf.}\ }\textbf {\bibinfo {volume} {10}},\ \bibinfo {pages} {46} (\bibinfo {year} {2024}{\natexlab{b}})},\ \Eprint {http://arxiv.org/abs/arXiv:2308.03054} {arXiv:2308.03054} \BibitemShut {NoStop}%
\bibitem [{NHJ()}]{NHJosephsonSM}%
  \BibitemOpen
  \href@noop {} {}\bibinfo {note} {See the Appendix~\ref{sec:SI} for the derivation of the magnonic Josephson equations, the Appendix~\ref{sec:SIV} for the derivation of the equation for $ \varphi$ as a function of the magnon-magnon interaction, the Appendix~\ref{sec:SII} for spin dynamics with different pumping rates, and the Appendix~\ref{sec:SIII} for the parameter dependence of our results.}\BibitemShut {Stop}%
\bibitem [{\citenamefont {Josephson}(1962)}]{Josephson1962}%
  \BibitemOpen
  \bibfield  {author} {\bibinfo {author} {\bibfnamefont {B.~D.}\ \bibnamefont {Josephson}},\ }\bibfield  {title} {\enquote {\bibinfo {title} {Possible new effects in superconductive tunnelling},}\ }\href {https://doi.org/10.1016/0031-9163(62)91369-0} {\bibfield  {journal} {\bibinfo  {journal} {Phys. Lett.}\ }\textbf {\bibinfo {volume} {1}},\ \bibinfo {pages} {251} (\bibinfo {year} {1962})}\BibitemShut {NoStop}%
\bibitem [{\citenamefont {Nakata}\ \emph {et~al.}(2015{\natexlab{b}})\citenamefont {Nakata}, \citenamefont {Simon},\ and\ \citenamefont {Loss}}]{magnonWF}%
  \BibitemOpen
  \bibfield  {author} {\bibinfo {author} {\bibfnamefont {K.}~\bibnamefont {Nakata}}, \bibinfo {author} {\bibfnamefont {P.}~\bibnamefont {Simon}}, \ and\ \bibinfo {author} {\bibfnamefont {D.}~\bibnamefont {Loss}},\ }\bibfield  {title} {\enquote {\bibinfo {title} {Wiedemann-{F}ranz law for magnon transport},}\ }\href {https://doi.org/10.1103/PhysRevB.92.134425} {\bibfield  {journal} {\bibinfo  {journal} {Phys. Rev. B}\ }\textbf {\bibinfo {volume} {92}},\ \bibinfo {pages} {134425} (\bibinfo {year} {2015}{\natexlab{b}})},\ \Eprint {http://arxiv.org/abs/arXiv:1507.03807} {arXiv:1507.03807} \BibitemShut {NoStop}%
\bibitem [{\citenamefont {Nakata}\ \emph {et~al.}(2017)\citenamefont {Nakata}, \citenamefont {Simon},\ and\ \citenamefont {Loss}}]{ReviewMagnon}%
  \BibitemOpen
  \bibfield  {author} {\bibinfo {author} {\bibfnamefont {K.}~\bibnamefont {Nakata}}, \bibinfo {author} {\bibfnamefont {P.}~\bibnamefont {Simon}}, \ and\ \bibinfo {author} {\bibfnamefont {D.}~\bibnamefont {Loss}},\ }\bibfield  {title} {\enquote {\bibinfo {title} {Spin currents and magnon dynamics in insulating magnets},}\ }\href {https://doi.org/10.1088/1361-6463/aa5b09} {\bibfield  {journal} {\bibinfo  {journal} {J. Phys. D: Appl. Phys.}\ }\textbf {\bibinfo {volume} {50}},\ \bibinfo {pages} {114004} (\bibinfo {year} {2017})},\ \Eprint {http://arxiv.org/abs/arXiv:1610.08901} {arXiv:1610.08901} \BibitemShut {NoStop}%
\bibitem [{\citenamefont {Zhang}\ \emph {et~al.}(2022)\citenamefont {Zhang}, \citenamefont {Gu}, \citenamefont {Li}, \citenamefont {Hu},\ and\ \citenamefont {Jiang}}]{JDtheory}%
  \BibitemOpen
  \bibfield  {author} {\bibinfo {author} {\bibfnamefont {Y.}~\bibnamefont {Zhang}}, \bibinfo {author} {\bibfnamefont {Y.}~\bibnamefont {Gu}}, \bibinfo {author} {\bibfnamefont {P.}~\bibnamefont {Li}}, \bibinfo {author} {\bibfnamefont {J.}~\bibnamefont {Hu}}, \ and\ \bibinfo {author} {\bibfnamefont {K.}~\bibnamefont {Jiang}},\ }\bibfield  {title} {\enquote {\bibinfo {title} {General theory of {J}osephson diodes},}\ }\href {https://doi.org/10.1103/PhysRevX.12.041013} {\bibfield  {journal} {\bibinfo  {journal} {Phys. Rev. X}\ }\textbf {\bibinfo {volume} {12}},\ \bibinfo {pages} {041013} (\bibinfo {year} {2022})},\ \Eprint {http://arxiv.org/abs/arXiv:2112.08901} {arXiv:2112.08901} \BibitemShut {NoStop}%
\bibitem [{\citenamefont {Davydova}\ \emph {et~al.}(2022)\citenamefont {Davydova}, \citenamefont {Prembabu},\ and\ \citenamefont {Fu}}]{JDtheoryLiangFu}%
  \BibitemOpen
  \bibfield  {author} {\bibinfo {author} {\bibfnamefont {M.}~\bibnamefont {Davydova}}, \bibinfo {author} {\bibfnamefont {S.}~\bibnamefont {Prembabu}}, \ and\ \bibinfo {author} {\bibfnamefont {L.}~\bibnamefont {Fu}},\ }\bibfield  {title} {\enquote {\bibinfo {title} {Universal {J}osephson diode effect},}\ }\href {https://www.science.org/doi/full/10.1126/sciadv.abo0309} {\bibfield  {journal} {\bibinfo  {journal} {Sci. Adv.}\ }\textbf {\bibinfo {volume} {8}},\ \bibinfo {pages} {eabo0309} (\bibinfo {year} {2022})},\ \Eprint {http://arxiv.org/abs/arXiv:2201.00831} {arXiv:2201.00831} \BibitemShut {NoStop}%
\bibitem [{\citenamefont {Wu}\ \emph {et~al.}(2022)\citenamefont {Wu}, \citenamefont {Wang}, \citenamefont {Xu}, \citenamefont {Sivakumar}, \citenamefont {Pasco}, \citenamefont {Filippozzi}, \citenamefont {Parkin}, \citenamefont {Zeng}, \citenamefont {McQueen},\ and\ \citenamefont {Ali}}]{JDexp}%
  \BibitemOpen
  \bibfield  {author} {\bibinfo {author} {\bibfnamefont {H.}~\bibnamefont {Wu}}, \bibinfo {author} {\bibfnamefont {Y.}~\bibnamefont {Wang}}, \bibinfo {author} {\bibfnamefont {Y.}~\bibnamefont {Xu}}, \bibinfo {author} {\bibfnamefont {P.~K.}\ \bibnamefont {Sivakumar}}, \bibinfo {author} {\bibfnamefont {C.}~\bibnamefont {Pasco}}, \bibinfo {author} {\bibfnamefont {U.}~\bibnamefont {Filippozzi}}, \bibinfo {author} {\bibfnamefont {S.~S.~P.}\ \bibnamefont {Parkin}}, \bibinfo {author} {\bibfnamefont {Y.-J.}\ \bibnamefont {Zeng}}, \bibinfo {author} {\bibfnamefont {T.}~\bibnamefont {McQueen}}, \ and\ \bibinfo {author} {\bibfnamefont {M.~N.}\ \bibnamefont {Ali}},\ }\bibfield  {title} {\enquote {\bibinfo {title} {The field-free {J}osephson diode in a van der {W}aals heterostructure},}\ }\href {https://doi.org/10.1038/s41586-022-04504-8} {\bibfield  {journal} {\bibinfo  {journal} {Nature}\ }\textbf {\bibinfo {volume} {604}},\ \bibinfo {pages} {653} (\bibinfo {year} {2022})},\ \Eprint {http://arxiv.org/abs/arXiv:2103.15809} {arXiv:2103.15809} \BibitemShut {NoStop}%
\bibitem [{\citenamefont {Parkin}\ \emph {et~al.}(1990)\citenamefont {Parkin}, \citenamefont {More},\ and\ \citenamefont {Roche}}]{ParkinPRL1990}%
  \BibitemOpen
  \bibfield  {author} {\bibinfo {author} {\bibfnamefont {S.~S.~P.}\ \bibnamefont {Parkin}}, \bibinfo {author} {\bibfnamefont {N.}~\bibnamefont {More}}, \ and\ \bibinfo {author} {\bibfnamefont {K.~P.}\ \bibnamefont {Roche}},\ }\bibfield  {title} {\enquote {\bibinfo {title} {Oscillations in exchange coupling and magnetoresistance in metallic superlattice structures: {C}o/{R}u, {C}o/{C}r, and {F}e/{C}r},}\ }\href {https://doi.org/10.1103/PhysRevLett.64.2304} {\bibfield  {journal} {\bibinfo  {journal} {Phys. Rev. Lett.}\ }\textbf {\bibinfo {volume} {64}},\ \bibinfo {pages} {2304} (\bibinfo {year} {1990})}\BibitemShut {NoStop}%
\bibitem [{\citenamefont {Parkin}\ \emph {et~al.}(1991)\citenamefont {Parkin}, \citenamefont {Bhadra},\ and\ \citenamefont {Roche}}]{ParkinPRL1991}%
  \BibitemOpen
  \bibfield  {author} {\bibinfo {author} {\bibfnamefont {S.~S.~P.}\ \bibnamefont {Parkin}}, \bibinfo {author} {\bibfnamefont {R.}~\bibnamefont {Bhadra}}, \ and\ \bibinfo {author} {\bibfnamefont {K.~P.}\ \bibnamefont {Roche}},\ }\bibfield  {title} {\enquote {\bibinfo {title} {Oscillatory magnetic exchange coupling through thin copper layers},}\ }\href {https://doi.org/10.1103/PhysRevLett.66.2152} {\bibfield  {journal} {\bibinfo  {journal} {Phys. Rev. Lett.}\ }\textbf {\bibinfo {volume} {66}},\ \bibinfo {pages} {2152} (\bibinfo {year} {1991})}\BibitemShut {NoStop}%
\bibitem [{\citenamefont {Husain}\ \emph {et~al.}(2022)\citenamefont {Husain}, \citenamefont {Pal}, \citenamefont {Chen}, \citenamefont {Kumar}, \citenamefont {Kumar}, \citenamefont {Mondal}, \citenamefont {Behera}, \citenamefont {Gupta}, \citenamefont {Hait}, \citenamefont {Gupta}, \citenamefont {Brucas}, \citenamefont {Sanyal}, \citenamefont {Barman}, \citenamefont {Chaudhary},\ and\ \citenamefont {Svedlindh}}]{HusainPRB2022}%
  \BibitemOpen
  \bibfield  {author} {\bibinfo {author} {\bibfnamefont {S.}~\bibnamefont {Husain}}, \bibinfo {author} {\bibfnamefont {S.}~\bibnamefont {Pal}}, \bibinfo {author} {\bibfnamefont {X.}~\bibnamefont {Chen}}, \bibinfo {author} {\bibfnamefont {P.}~\bibnamefont {Kumar}}, \bibinfo {author} {\bibfnamefont {A.}~\bibnamefont {Kumar}}, \bibinfo {author} {\bibfnamefont {A.~K.}\ \bibnamefont {Mondal}}, \bibinfo {author} {\bibfnamefont {N.}~\bibnamefont {Behera}}, \bibinfo {author} {\bibfnamefont {N.~K.}\ \bibnamefont {Gupta}}, \bibinfo {author} {\bibfnamefont {S.}~\bibnamefont {Hait}}, \bibinfo {author} {\bibfnamefont {R.}~\bibnamefont {Gupta}}, \bibinfo {author} {\bibfnamefont {R.}~\bibnamefont {Brucas}}, \bibinfo {author} {\bibfnamefont {B.}~\bibnamefont {Sanyal}}, \bibinfo {author} {\bibfnamefont {A.}~\bibnamefont {Barman}}, \bibinfo {author} {\bibfnamefont {S.}~\bibnamefont {Chaudhary}}, \ and\ \bibinfo {author} {\bibfnamefont {P.}~\bibnamefont {Svedlindh}},\ }\bibfield  {title} {\enquote {\bibinfo {title} {Large {D}zyaloshinskii-{M}oriya interaction and atomic layer thickness dependence in a ferromagnet-${{\mathrm{WS}}}_{2}$ heterostructure},}\ }\href {https://doi.org/10.1103/PhysRevB.105.064422} {\bibfield  {journal} {\bibinfo  {journal} {Phys. Rev. B}\ }\textbf {\bibinfo {volume} {105}},\ \bibinfo {pages} {064422} (\bibinfo {year} {2022})}\BibitemShut {NoStop}%
\bibitem [{\citenamefont {Yun}\ \emph {et~al.}(2023)\citenamefont {Yun}, \citenamefont {Cui}, \citenamefont {Cui}, \citenamefont {He}, \citenamefont {Chang}, \citenamefont {Zhu}, \citenamefont {Yan}, \citenamefont {Guo}, \citenamefont {Xie}, \citenamefont {Zhang} \emph {et~al.}}]{yun2023anisotropic}%
  \BibitemOpen
  \bibfield  {author} {\bibinfo {author} {\bibfnamefont {J.}~\bibnamefont {Yun}}, \bibinfo {author} {\bibfnamefont {B.}~\bibnamefont {Cui}}, \bibinfo {author} {\bibfnamefont {Q.}~\bibnamefont {Cui}}, \bibinfo {author} {\bibfnamefont {X.}~\bibnamefont {He}}, \bibinfo {author} {\bibfnamefont {Y.}~\bibnamefont {Chang}}, \bibinfo {author} {\bibfnamefont {Y.}~\bibnamefont {Zhu}}, \bibinfo {author} {\bibfnamefont {Z.}~\bibnamefont {Yan}}, \bibinfo {author} {\bibfnamefont {X.}~\bibnamefont {Guo}}, \bibinfo {author} {\bibfnamefont {H.}~\bibnamefont {Xie}}, \bibinfo {author} {\bibfnamefont {J.}~\bibnamefont {Zhang}},  \emph {et~al.},\ }\bibfield  {title} {\enquote {\bibinfo {title} {Anisotropic interlayer {D}zyaloshinskii--{M}oriya interaction in synthetic ferromagnetic/antiferromagnetic sandwiches},}\ }\href {https://doi.org/10.1002/adfm.202301731} {\bibfield  {journal} {\bibinfo  {journal} {Adv. Funct. Mater.}\ ,\ \bibinfo {pages} {2301731}} (\bibinfo {year} {2023})}\BibitemShut {NoStop}%
\bibitem [{\citenamefont {Srivastava}\ \emph {et~al.}(2018)\citenamefont {Srivastava}, \citenamefont {Schott}, \citenamefont {Juge}, \citenamefont {Krizakova}, \citenamefont {Belmeguenai}, \citenamefont {Roussign{\'e}}, \citenamefont {Bernand-Mantel}, \citenamefont {Ranno}, \citenamefont {Pizzini}, \citenamefont {Ch{\'e}rif} \emph {et~al.}}]{srivastava2018large}%
  \BibitemOpen
  \bibfield  {author} {\bibinfo {author} {\bibfnamefont {T.}~\bibnamefont {Srivastava}}, \bibinfo {author} {\bibfnamefont {M.}~\bibnamefont {Schott}}, \bibinfo {author} {\bibfnamefont {R.}~\bibnamefont {Juge}}, \bibinfo {author} {\bibfnamefont {V.}~\bibnamefont {Krizakova}}, \bibinfo {author} {\bibfnamefont {M.}~\bibnamefont {Belmeguenai}}, \bibinfo {author} {\bibfnamefont {Y.}~\bibnamefont {Roussign{\'e}}}, \bibinfo {author} {\bibfnamefont {A.}~\bibnamefont {Bernand-Mantel}}, \bibinfo {author} {\bibfnamefont {L.}~\bibnamefont {Ranno}}, \bibinfo {author} {\bibfnamefont {S.}~\bibnamefont {Pizzini}}, \bibinfo {author} {\bibfnamefont {S.-M.}\ \bibnamefont {Ch{\'e}rif}},  \emph {et~al.},\ }\bibfield  {title} {\enquote {\bibinfo {title} {Large-voltage tuning of {D}zyaloshinskii--{M}oriya interactions: A route toward dynamic control of skyrmion chirality},}\ }\href {https://doi.org/10.1021/acs.nanolett.8b01502} {\bibfield  {journal} {\bibinfo  {journal} {Nano Lett.}\ }\textbf {\bibinfo {volume} {18}},\ \bibinfo {pages} {4871} (\bibinfo {year} {2018})}\BibitemShut {NoStop}%
\bibitem [{\citenamefont {Fillion}\ \emph {et~al.}(2022)\citenamefont {Fillion}, \citenamefont {Fischer}, \citenamefont {Kumar}, \citenamefont {Fassatoui}, \citenamefont {Pizzini}, \citenamefont {Ranno}, \citenamefont {Ourdani}, \citenamefont {Belmeguenai}, \citenamefont {Roussign{\'e}}, \citenamefont {Ch{\'e}rif} \emph {et~al.}}]{fillion2022gate}%
  \BibitemOpen
  \bibfield  {author} {\bibinfo {author} {\bibfnamefont {C.-E.}\ \bibnamefont {Fillion}}, \bibinfo {author} {\bibfnamefont {J.}~\bibnamefont {Fischer}}, \bibinfo {author} {\bibfnamefont {R.}~\bibnamefont {Kumar}}, \bibinfo {author} {\bibfnamefont {A.}~\bibnamefont {Fassatoui}}, \bibinfo {author} {\bibfnamefont {S.}~\bibnamefont {Pizzini}}, \bibinfo {author} {\bibfnamefont {L.}~\bibnamefont {Ranno}}, \bibinfo {author} {\bibfnamefont {D.}~\bibnamefont {Ourdani}}, \bibinfo {author} {\bibfnamefont {M.}~\bibnamefont {Belmeguenai}}, \bibinfo {author} {\bibfnamefont {Y.}~\bibnamefont {Roussign{\'e}}}, \bibinfo {author} {\bibfnamefont {S.-M.}\ \bibnamefont {Ch{\'e}rif}},  \emph {et~al.},\ }\bibfield  {title} {\enquote {\bibinfo {title} {Gate-controlled skyrmion and domain wall chirality},}\ }\href {https://doi.org/10.1038/s41467-022-32959-w} {\bibfield  {journal} {\bibinfo  {journal} {Nat. Commun.}\ }\textbf {\bibinfo {volume} {13}},\ \bibinfo {pages} {5257} (\bibinfo {year} {2022})},\ \Eprint {http://arxiv.org/abs/arXiv:2204.04031} {arXiv:2204.04031} \BibitemShut {NoStop}%
\bibitem [{\citenamefont {Koyama}\ \emph {et~al.}(2018)\citenamefont {Koyama}, \citenamefont {Nakatani}, \citenamefont {Ieda},\ and\ \citenamefont {Chiba}}]{koyama2018electric}%
  \BibitemOpen
  \bibfield  {author} {\bibinfo {author} {\bibfnamefont {T.}~\bibnamefont {Koyama}}, \bibinfo {author} {\bibfnamefont {Y.}~\bibnamefont {Nakatani}}, \bibinfo {author} {\bibfnamefont {J.}~\bibnamefont {Ieda}}, \ and\ \bibinfo {author} {\bibfnamefont {D.}~\bibnamefont {Chiba}},\ }\bibfield  {title} {\enquote {\bibinfo {title} {Electric field control of magnetic domain wall motion via modulation of the {D}zyaloshinskii-{M}oriya interaction},}\ }\href {https://doi.org/10.1126/sciadv.aav0265} {\bibfield  {journal} {\bibinfo  {journal} {Sci. Adv.}\ }\textbf {\bibinfo {volume} {4}},\ \bibinfo {pages} {eaav0265} (\bibinfo {year} {2018})}\BibitemShut {NoStop}%
\bibitem [{\citenamefont {Vedmedenko}\ \emph {et~al.}(2019)\citenamefont {Vedmedenko}, \citenamefont {Riego}, \citenamefont {Arregi},\ and\ \citenamefont {Berger}}]{PRL2019Vedmedenko}%
  \BibitemOpen
  \bibfield  {author} {\bibinfo {author} {\bibfnamefont {E.~Y.}\ \bibnamefont {Vedmedenko}}, \bibinfo {author} {\bibfnamefont {P.}~\bibnamefont {Riego}}, \bibinfo {author} {\bibfnamefont {J.~A.}\ \bibnamefont {Arregi}}, \ and\ \bibinfo {author} {\bibfnamefont {A.}~\bibnamefont {Berger}},\ }\bibfield  {title} {\enquote {\bibinfo {title} {Interlayer {D}zyaloshinskii-{M}oriya interactions},}\ }\href {https://doi.org/10.1103/PhysRevLett.122.257202} {\bibfield  {journal} {\bibinfo  {journal} {Phys. Rev. Lett.}\ }\textbf {\bibinfo {volume} {122}},\ \bibinfo {pages} {257202} (\bibinfo {year} {2019})},\ \Eprint {http://arxiv.org/abs/arXiv:1803.10570} {arXiv:1803.10570} \BibitemShut {NoStop}%
\bibitem [{\citenamefont {Cornelissen}\ \emph {et~al.}(2015)\citenamefont {Cornelissen}, \citenamefont {Liu}, \citenamefont {Duine}, \citenamefont {Youssef},\ and\ \citenamefont {v.~Wees}}]{LongMagnon2015NatPhys}%
  \BibitemOpen
  \bibfield  {author} {\bibinfo {author} {\bibfnamefont {L.~J.}\ \bibnamefont {Cornelissen}}, \bibinfo {author} {\bibfnamefont {J.}~\bibnamefont {Liu}}, \bibinfo {author} {\bibfnamefont {R.~A.}\ \bibnamefont {Duine}}, \bibinfo {author} {\bibfnamefont {J.~B.}\ \bibnamefont {Youssef}}, \ and\ \bibinfo {author} {\bibfnamefont {B.~J.}\ \bibnamefont {v.~Wees}},\ }\bibfield  {title} {\enquote {\bibinfo {title} {Long-distance transport of magnon spin information in a magnetic insulator at room temperature},}\ }\href {https://doi.org/10.1038/nphys3465} {\bibfield  {journal} {\bibinfo  {journal} {Nat. Phys.}\ }\textbf {\bibinfo {volume} {11}},\ \bibinfo {pages} {1022} (\bibinfo {year} {2015})},\ \Eprint {http://arxiv.org/abs/arXiv:1505.06325} {arXiv:1505.06325} \BibitemShut {NoStop}%
\bibitem [{\citenamefont {Boltyk}\ \emph {et~al.}(2012)\citenamefont {Boltyk}, \citenamefont {Dzyapko}, \citenamefont {Demidov}, \citenamefont {Berloff},\ and\ \citenamefont {Demokritov}}]{MagnonCoherenceLength}%
  \BibitemOpen
  \bibfield  {author} {\bibinfo {author} {\bibfnamefont {P.~N.}\ \bibnamefont {Boltyk}}, \bibinfo {author} {\bibfnamefont {O.}~\bibnamefont {Dzyapko}}, \bibinfo {author} {\bibfnamefont {V.~E.}\ \bibnamefont {Demidov}}, \bibinfo {author} {\bibfnamefont {N.~G.}\ \bibnamefont {Berloff}}, \ and\ \bibinfo {author} {\bibfnamefont {S.~O.}\ \bibnamefont {Demokritov}},\ }\bibfield  {title} {\enquote {\bibinfo {title} {Spatially non-uniform ground state and quantized vortices in a two-component {B}ose-{E}instein condensate of magnons},}\ }\href {https://doi.org/10.1038/srep00482} {\bibfield  {journal} {\bibinfo  {journal} {Sci. Rep.}\ }\textbf {\bibinfo {volume} {2}},\ \bibinfo {pages} {482} (\bibinfo {year} {2012})}\BibitemShut {NoStop}%
\bibitem [{\citenamefont {Pirro}\ \emph {et~al.}(2021)\citenamefont {Pirro}, \citenamefont {Vasyuchka}, \citenamefont {Serga},\ and\ \citenamefont {Hillebrands}}]{ReviewCoherentMagnon2021}%
  \BibitemOpen
  \bibfield  {author} {\bibinfo {author} {\bibfnamefont {P.}~\bibnamefont {Pirro}}, \bibinfo {author} {\bibfnamefont {V.~I.}\ \bibnamefont {Vasyuchka}}, \bibinfo {author} {\bibfnamefont {A.~A.}\ \bibnamefont {Serga}}, \ and\ \bibinfo {author} {\bibfnamefont {B.}~\bibnamefont {Hillebrands}},\ }\bibfield  {title} {\enquote {\bibinfo {title} {Advances in coherent magnonics},}\ }\href {https://doi.org/10.1038/s41578-021-00332-w} {\bibfield  {journal} {\bibinfo  {journal} {Nat. Rev. Mater.}\ }\textbf {\bibinfo {volume} {6}},\ \bibinfo {pages} {1114} (\bibinfo {year} {2021})}\BibitemShut {NoStop}%
\bibitem [{\citenamefont {Kreil}\ \emph {et~al.}(2021)\citenamefont {Kreil}, \citenamefont {Shmarova}, \citenamefont {Frey}, \citenamefont {Pomyalov}, \citenamefont {L'vov}, \citenamefont {Melkov}, \citenamefont {Serga},\ and\ \citenamefont {Hillebrands}}]{ExpMagnonJosephson}%
  \BibitemOpen
  \bibfield  {author} {\bibinfo {author} {\bibfnamefont {A.~J.~E.}\ \bibnamefont {Kreil}}, \bibinfo {author} {\bibfnamefont {H.~Y.~M.}\ \bibnamefont {Shmarova}}, \bibinfo {author} {\bibfnamefont {P.}~\bibnamefont {Frey}}, \bibinfo {author} {\bibfnamefont {A.}~\bibnamefont {Pomyalov}}, \bibinfo {author} {\bibfnamefont {V.~S.}\ \bibnamefont {L'vov}}, \bibinfo {author} {\bibfnamefont {G.~A.}\ \bibnamefont {Melkov}}, \bibinfo {author} {\bibfnamefont {A.~A.}\ \bibnamefont {Serga}}, \ and\ \bibinfo {author} {\bibfnamefont {B.}~\bibnamefont {Hillebrands}},\ }\bibfield  {title} {\enquote {\bibinfo {title} {Experimental observation of {J}osephson oscillations in a room-temperature {B}ose-{E}instein magnon condensate},}\ }\href {https://doi.org/10.1103/PhysRevB.104.144414} {\bibfield  {journal} {\bibinfo  {journal} {Phys. Rev. B}\ }\textbf {\bibinfo {volume} {104}},\ \bibinfo {pages} {144414} (\bibinfo {year} {2021})},\ \Eprint {http://arxiv.org/abs/arXiv:1911.07802} {arXiv:1911.07802} \BibitemShut {NoStop}%
\bibitem [{\citenamefont {Bozhko}\ \emph {et~al.}(2016)\citenamefont {Bozhko}, \citenamefont {Serga}, \citenamefont {Clausen}, \citenamefont {Vasyuchka}, \citenamefont {Heussner}, \citenamefont {Melkov}, \citenamefont {Pomyalov}, \citenamefont {L'vov},\ and\ \citenamefont {Hillebrands}}]{MagnonSupercurrent}%
  \BibitemOpen
  \bibfield  {author} {\bibinfo {author} {\bibfnamefont {D.~A.}\ \bibnamefont {Bozhko}}, \bibinfo {author} {\bibfnamefont {A.~A.}\ \bibnamefont {Serga}}, \bibinfo {author} {\bibfnamefont {P.}~\bibnamefont {Clausen}}, \bibinfo {author} {\bibfnamefont {V.~I.}\ \bibnamefont {Vasyuchka}}, \bibinfo {author} {\bibfnamefont {F.}~\bibnamefont {Heussner}}, \bibinfo {author} {\bibfnamefont {G.~A.}\ \bibnamefont {Melkov}}, \bibinfo {author} {\bibfnamefont {A.}~\bibnamefont {Pomyalov}}, \bibinfo {author} {\bibfnamefont {V.~S.}\ \bibnamefont {L'vov}}, \ and\ \bibinfo {author} {\bibfnamefont {B.}~\bibnamefont {Hillebrands}},\ }\bibfield  {title} {\enquote {\bibinfo {title} {Supercurrent in a room-temperature {B}ose-{E}instein magnon condensate},}\ }\href {https://doi.org/10.1038/nphys3838} {\bibfield  {journal} {\bibinfo  {journal} {Nat. Phys.}\ }\textbf {\bibinfo {volume} {12}},\ \bibinfo {pages} {1057} (\bibinfo {year} {2016})},\ \Eprint {http://arxiv.org/abs/arXiv:1503.00482} {arXiv:1503.00482} \BibitemShut {NoStop}%
\bibitem [{\citenamefont {Heinrich}\ \emph {et~al.}(2011)\citenamefont {Heinrich}, \citenamefont {Burrowes}, \citenamefont {Montoya}, \citenamefont {Kardasz}, \citenamefont {Girt}, \citenamefont {Song}, \citenamefont {Sun},\ and\ \citenamefont {Wu}}]{YIGdamping2011}%
  \BibitemOpen
  \bibfield  {author} {\bibinfo {author} {\bibfnamefont {B.}~\bibnamefont {Heinrich}}, \bibinfo {author} {\bibfnamefont {C.}~\bibnamefont {Burrowes}}, \bibinfo {author} {\bibfnamefont {E.}~\bibnamefont {Montoya}}, \bibinfo {author} {\bibfnamefont {B.}~\bibnamefont {Kardasz}}, \bibinfo {author} {\bibfnamefont {E.}~\bibnamefont {Girt}}, \bibinfo {author} {\bibfnamefont {Y.~Y.}\ \bibnamefont {Song}}, \bibinfo {author} {\bibfnamefont {Y.}~\bibnamefont {Sun}}, \ and\ \bibinfo {author} {\bibfnamefont {M.}~\bibnamefont {Wu}},\ }\bibfield  {title} {\enquote {\bibinfo {title} {Spin pumping at the magnetic insulator ({YIG})/normal metal ({A}u) interfaces},}\ }\href {https://doi.org/10.1103/PhysRevLett.107.066604} {\bibfield  {journal} {\bibinfo  {journal} {Phys. Rev. Lett.}\ }\textbf {\bibinfo {volume} {107}},\ \bibinfo {pages} {066604} (\bibinfo {year} {2011})}\BibitemShut {NoStop}%
\bibitem [{\citenamefont {Margineda}\ \emph {et~al.}(2023)\citenamefont {Margineda}, \citenamefont {Claydon}, \citenamefont {Qejvanaj},\ and\ \citenamefont {Checkley}}]{AJE_exp_2023}%
  \BibitemOpen
  \bibfield  {author} {\bibinfo {author} {\bibfnamefont {D.}~\bibnamefont {Margineda}}, \bibinfo {author} {\bibfnamefont {J.~S.}\ \bibnamefont {Claydon}}, \bibinfo {author} {\bibfnamefont {F.}~\bibnamefont {Qejvanaj}}, \ and\ \bibinfo {author} {\bibfnamefont {C.}~\bibnamefont {Checkley}},\ }\bibfield  {title} {\enquote {\bibinfo {title} {Observation of anomalous {J}osephson effect in nonequilibrium {A}ndreev interferometers},}\ }\href {https://doi.org/10.1103/PhysRevB.107.L100502} {\bibfield  {journal} {\bibinfo  {journal} {Phys. Rev. B}\ }\textbf {\bibinfo {volume} {107}},\ \bibinfo {pages} {L100502} (\bibinfo {year} {2023})},\ \Eprint {http://arxiv.org/abs/arXiv:2105.13968} {arXiv:2105.13968} \BibitemShut {NoStop}%
\bibitem [{\citenamefont {Dolgirev}\ \emph {et~al.}(2019{\natexlab{a}})\citenamefont {Dolgirev}, \citenamefont {Kalenkov},\ and\ \citenamefont {Zaikin}}]{AJE_Theory_2019}%
  \BibitemOpen
  \bibfield  {author} {\bibinfo {author} {\bibfnamefont {P.~E.}\ \bibnamefont {Dolgirev}}, \bibinfo {author} {\bibfnamefont {M.~S.}\ \bibnamefont {Kalenkov}}, \ and\ \bibinfo {author} {\bibfnamefont {A.~D.}\ \bibnamefont {Zaikin}},\ }\bibfield  {title} {\enquote {\bibinfo {title} {Interplay between {J}osephson and {A}haronov-{B}ohm effects in {A}ndreev interferometers},}\ }\href {https://doi.org/10.1038/s41598-018-37653-w} {\bibfield  {journal} {\bibinfo  {journal} {Sci. Rep.}\ }\textbf {\bibinfo {volume} {9}},\ \bibinfo {pages} {1301} (\bibinfo {year} {2019}{\natexlab{a}})},\ \Eprint {http://arxiv.org/abs/arXiv:1807.10339} {arXiv:1807.10339} \BibitemShut {NoStop}%
\bibitem [{\citenamefont {Dolgirev}\ \emph {et~al.}(2019{\natexlab{b}})\citenamefont {Dolgirev}, \citenamefont {Kalenkov}, \citenamefont {Tarkhov},\ and\ \citenamefont {Zaikin}}]{AJE_Theory_2019PRB}%
  \BibitemOpen
  \bibfield  {author} {\bibinfo {author} {\bibfnamefont {P.~E.}\ \bibnamefont {Dolgirev}}, \bibinfo {author} {\bibfnamefont {M.~S.}\ \bibnamefont {Kalenkov}}, \bibinfo {author} {\bibfnamefont {A.~E.}\ \bibnamefont {Tarkhov}}, \ and\ \bibinfo {author} {\bibfnamefont {A.~D.}\ \bibnamefont {Zaikin}},\ }\bibfield  {title} {\enquote {\bibinfo {title} {Phase-coherent electron transport in asymmetric crosslike {A}ndreev interferometers},}\ }\href {https://doi.org/10.1103/PhysRevB.100.054511} {\bibfield  {journal} {\bibinfo  {journal} {Phys. Rev. B}\ }\textbf {\bibinfo {volume} {100}},\ \bibinfo {pages} {054511} (\bibinfo {year} {2019}{\natexlab{b}})},\ \Eprint {http://arxiv.org/abs/arXiv:1906.07305} {arXiv:1906.07305} \BibitemShut {NoStop}%
\bibitem [{\citenamefont {Hijano}\ \emph {et~al.}(2021)\citenamefont {Hijano}, \citenamefont {Ili\'{c}},\ and\ \citenamefont {Bergeret}}]{AJE_Theory_2021}%
  \BibitemOpen
  \bibfield  {author} {\bibinfo {author} {\bibfnamefont {A.}~\bibnamefont {Hijano}}, \bibinfo {author} {\bibfnamefont {S.}~\bibnamefont {Ili\'{c}}}, \ and\ \bibinfo {author} {\bibfnamefont {F.~S.}\ \bibnamefont {Bergeret}},\ }\bibfield  {title} {\enquote {\bibinfo {title} {Anomalous {A}ndreev interferometer: Study of an anomalous {J}osephson junction coupled to a normal wire},}\ }\href {https://doi.org/10.1103/PhysRevB.104.214515} {\bibfield  {journal} {\bibinfo  {journal} {Phys. Rev. B}\ }\textbf {\bibinfo {volume} {104}},\ \bibinfo {pages} {214515} (\bibinfo {year} {2021})},\ \Eprint {http://arxiv.org/abs/arXiv:2106.14021} {arXiv:2106.14021} \BibitemShut {NoStop}%
\end{thebibliography}%

\end{document}